# Upscaling-based modified deep bed filtration model to match hyper-exponential retention

Nastaran Khazali[a], Gabriel Malgaresi[b], Ludmila Kuzmina[c], Yuri Osipov[d], Thomas Russell[a], Pavel Bedrikovetsky[a]

[a] School of Chemical Engineering, University of Adelaide, SA, Australia.

[b] Predico Software, QLD, Australia.

[c] HSE University, Moscow, Russia.

[d] Moscow State University of Civil Engineering, Moscow, Russia.

## Abstract

Modelling of colloidal and nano-suspension transport in porous media has garnered significant attention due to the prevalence of these processes in many engineering applications. A number of experimental studies have reported retention profiles after coreflooding that are hyper-exponential, a feature that the traditional models for deep bed filtration are unable to capture. The aim of this work is the development of a model for binary particle transport which can account for hyper-exponential retention profiles (HERPs). Averaging of the binary model results in a non-linear dependence of the capture rate on the suspended particle concentration. The averaged model demonstrates how nonlinear capture behaviour arises due to particle heterogeneity even when individual particle populations filter traditionally. The model establishes that nonlinear capture behaviour, including the prediction of HERPs, is possible even for dilute particle suspensions given that particle heterogeneity is significant. An exact solution of the colloidal transport equations is presented for any non-linear suspension function and any step-wise-constant injected concentration. The binary suspension function cannot be written explicitly, and so we present asymptotic expansions for several limiting cases where the suspension function can be expressed directly. The asymptotic expansions show good agreement with the binary model, and correctly capture the HERPs calculated with the exact solution. Several laboratory tests exhibiting HERPs have been matched by the traditional, binary, and asymptotic expansion models. The traditional model significantly deviates from the laboratory data due to its inability to capture the hyper-exponential behaviour, while both the binary and asymptotic models capture this behaviour well.

| Nomenclature | | | |
|---|---|---|---|
| $C$ | Suspended concentration distribution, $[L^{-4}]$ | $U$ | Darcy flow velocity, $[LT^{-1}]$ |
| $c$ | Total suspended particle concentration, $[L^{-3}]$ | $x$ | Coordinate, $[L]$ |
| $c^0$ | Injected particle concentration, $[L^{-3}]$ | Greek Letters | |
| $F$ | Individual suspension function for particles with size r, $[L^{-5}]$ | $\beta$ | Formation damage coefficient, [-] |
| $f$ | Upscaled suspension function, $[L^{-4}]$ | $\Delta p$ | Pressure drop across the core, $[ML^{-1}T^{-2}]$ |
| $h$ | Filtration function, [-] | $\varepsilon$ | Asymptotically small value, [-] |
| $J$ | Impedance, [-] | $\Theta_\lambda$ | Dimensionless difference between filtration coefficients, [-] |
| $k_0$ | Permeability, $[L^2]$ | $\lambda$ | Filtration coefficient, $[L^{-1}]$ |
| $L$ | Core length, $[L]$ | $\mu$ | Viscosity, $[ML^{-1}T^{-1}]$ |
| $p$ | Pressure, $[ML^{-1}T^{-2}]$ | $\phi$ | Porosity, [-] |
| $r$ | Particle radius, $[L]$ | Abbreviations | |
| $S$ | Retained concentration distribution, $[L^{-4}]$ | BTC | Breakthrough curve |
| $s$ | Total deposit concentration, $[L^{-3}]$ | DBF | Deep bed filtration |
| $s_m$ | Dimensionless concentration of the capture vacancies, $[L^{-3}]$ | HERP | Hyper-exponential retention profile |
| $t$ | Time, $[T]$ | PVI | Pore volume injected |

## 1. Introduction

Colloidal transport in porous media occurs in various areas of chemical, environmental, and petroleum engineering [1]–[7].The transport of colloidal particles can contribute to enhancing contaminant transport in the vadose zone [8]–[10] or using nano technologies [11]–[13] and particle capture can alter the properties of the porous media, increasing the difficulty of fluid injection [14] and production [15]. A schematic of particle transport and the capture mechanisms in porous media is presented in Fig. 1.

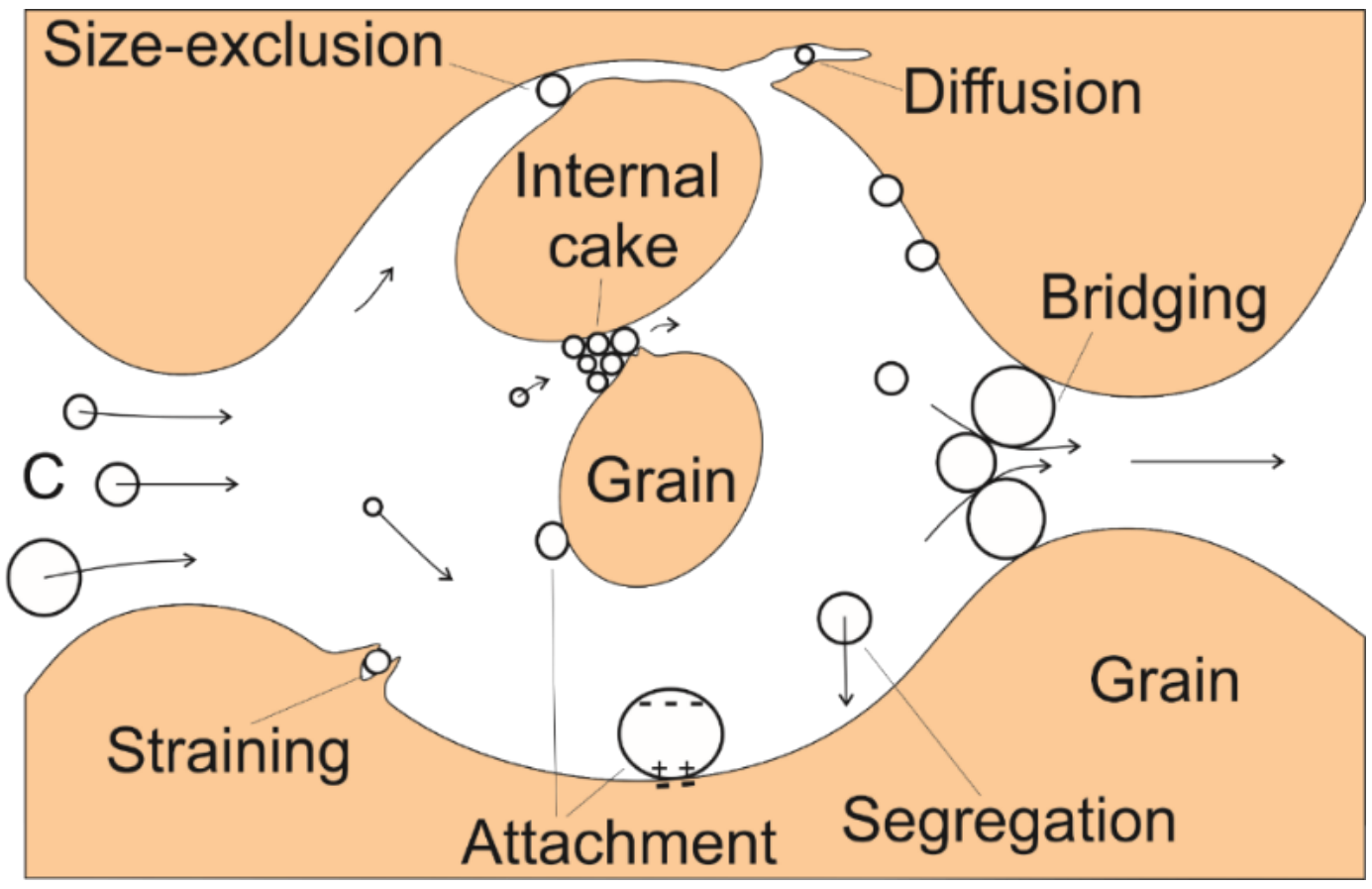

Fig. 1: Multiple capture mechanisms during colloidal and nano-suspension transport in porous media.

The governing system consists of the equations for mass balance for suspended and retained particles, and the particle capture rate [16]:

$$\phi \frac{\partial c}{\partial t} + U \frac{\partial c}{\partial x} = -\frac{\partial s}{\partial t} \qquad (1)$$

$$\frac{\partial s}{\partial t} = \lambda c U \qquad (2)$$

where $\phi$ is the rock porosity, $t$ is the time, $U$ is the fluid velocity, $x$ is the spatial dimension, $\lambda$ is the filtration coefficient, and $c$ and $s$ are the suspended and retained particle concentrations respectively.

The majority of applications utilize a constant filtration coefficient, $\lambda$. This assumption is based on the notion that the probability of particle capture is constant [17]. The exact solution for this system with constant particle injection is widely available [16], [18], and produces a retention profile that decreases in $x$ exponentially. Many laboratory studies report retention profiles that are not exponential. Profiles that are less sharp than an exponential profile are attributed to a decreasing capture probability with increasing retained particle concentration [19], or $\lambda(s)$. Many laboratory tests report a retention profile that is hyper-exponential; i.e. the retained concentration drops sharper than can be explained by an exponential retention profile [20]–[23]. Some authors suggest a depth-varying filtration coefficient to explain these profiles, and both outlet concentrations and retention profiles have been simultaneously fit with this model [24], [25]. The theoretical basis of this model is not well substantiated outside of good quality matching.

Some authors have claimed that particle interactions during capture, such as the formation of particle bridges at pore openings, could be the cause for retention hyper-exponentiality [26]. Having multiple mechanisms causing capture simultaneously has been suggested to be the reason for retention profiles not being exponential [20]–[22], [27]–[30]. Particle interactions during capture usually rely on assumptions of high suspended particle concentrations [31], similar to activity coefficients used in chemical reaction kinetics. However, some studies have reported hyper-exponential retention profiles even at small injected concentrations [32].

Hyper-exponential profiles have been explained by multiple populations of particles, whose standard exponential profiles, when summed, can result in hyper-exponentiality [17], [18], [20], [28], [29], [33]. Some authors have developed models to capture the effect of particle heterogeneity, but not all cases have successfully fit both breakthrough concentration and retention profile curves [34]–[36]. Existing models treat the different particle populations entirely separately, which prevents modelling a decrease in capture probability with increasing retention profiles [21], [22], [28]. Accounting for an $n$-particle system with distributed, but constant, filtration coefficients involves solving $n$ independent differential equations for the suspended concentrations, and *n* equations for the strained concentrations [29]. When the filtration coefficient decreases with the strained particle concentration, the *2n* equations are no longer independent and must be solved simultaneously which can induce prohibitive computational costs [29], [37]. Despite significant efforts in upscaling of colloidal transport [17], [38], [39], the averaged equations for multicomponent populations are not available.

In this work we remedy these shortfalls by modifying a recently developed model which proved that injection of multiple particle populations can be averaged exactly, and that the average particle capture rate is proportional not to *c* but to some function *f(c)*. These equations are modified to model 2-particle behaviour with equal occupation terms. While the resulting model captures this physical behaviour, the derived form for *f(c)* cannot be written explicitly, so in this work several asymptotic expansions are presented which allow for explicit presentation of the model. In addition, the averaged system of equations is solved allowing for the calculation of *c(x,t)* and *s(x,t)* for any function *f(c)*, Langmuir *λ(s)*, and stepwise varying injected concentration *c(0,t)*. For continuous injection tests, we match simultaneously the breakthrough suspended concentration, *c(1,t)*, and the hyper-exponential retained concentration profile, $s(x,t_0)$, which shows close agreement.

The structure of the paper is as follows. Section 2 introduces the mathematical model for particle transport with non-linear capture. Section 3 presents the expressions for the averaged suspended concentration and suspension function of binary colloidal transport system. Section 4 provides an analysis of the behaviour of diluted suspensions. Section 5 introduces asymptotic expressions for the suspension function. Section 6 presents the exact solution for 1D upscaled colloidal transport. Section 7 validates the exact solution with a presented numerical method for solving the system of governing equations. Section 8 includes treatment of laboratory data. Section 9 discusses the results and Section 10 concludes the paper.

## 2. Mathematical model for particle transport with non-linear capture

### 2.1 Assumptions of the model

We discuss single-phase flow of colloidal suspensions through a solid porous media accounting for injected particle heterogeneity. The assumptions of the model are as follows.

We assume one-dimensional incompressible flow. The suspended concentration is assumed to be sufficiently small such that the suspended colloids do not alter the density or viscosity of the fluid. The effect of the retained particles on the rock porosity is assumed to be negligible. Particle diffusion is neglected on the basis of sufficiently high velocities and sufficiently large particles. Each of the individual particle populations filters through the porous medium 'traditionally', i.e. with constant filtration coefficient and all particles occupy the same fraction of the available capture sites/area. Lastly the inverse of the rock permeability is assumed to vary linearly with the concentration of retained particles.

### 2.2 Upscaled governing system

Following the averaging procedure outlined in Appendix A, an ensemble of *n* populations, each with traditional capture rate, results in the following retention rate for the averaged retained particle concentration:

$$\frac{\partial s}{\partial t} = h(s)\, f(c)U \tag{3}$$

where *h(s)* is the filtration function, and *f(c)* is the suspension function.

The most common form of the filtration function is the Langmuir equation, commonly found in adsorption applications [40]:

$$h(s) = \left(1 - \frac{s}{s_m}\right) \tag{4}$$

where $s_m$ represents the total number of capture sites such that as $s \to s_m$, the capture rate tends to zero.

The modified form of Darcy's law accounting for a decrease in rock permeability due to particle retention is:

$$U = -\frac{k_0}{\mu(1+\beta s)}\frac{\partial p}{\partial x} \tag{5}$$

where $k_0$ is the initial rock permeability ($k_0=k(s=0)$), $\mu$ is the fluid viscosity, $\beta$ is the formation damage factor, and $p$ is the fluid pressure.

The initial and boundary conditions considered in this work correspond to zero suspended and retained particles in the system initially, and the injection of particles given by any function of time $c(0,t)=c^0(t)$:

$$t=0:\ c=s=0; \quad x=0:\ c=c^0(t) \tag{6}$$

The system can be non-dimensionalised by introducing the following non-dimensionless parameters:

$$x \to \frac{x}{L}, t \to \frac{Ut}{\phi L}, c \to \frac{c}{c^0}, s \to \frac{s}{\phi c^0}, p \to \frac{k_0 p}{UL\mu}, s_m \to \frac{s_m}{\phi c^0}, f(c) \to \frac{Lf(c)}{c^0} \tag{7}$$

where $L$ is the system length, and $c^0$ is the initial injected concentration ($c^0=c^0(0)$).

Using these new variables, the dimensionless system of governing equations becomes:

$$\frac{\partial c}{\partial t} + \frac{\partial c}{\partial x} = -\frac{\partial s}{\partial t} \tag{8}$$

$$\frac{\partial s}{\partial t} = h(s)\, f(c) \tag{9}$$

$$1 = -\frac{1}{1+\beta\phi c^0 s}\frac{\partial p}{\partial x} \tag{10}$$

The dimensionless initial and boundary conditions are:

$$t=0: c=s=0; \ x=0: c=\frac{c^0(t)}{c^0(0)} \tag{11}$$

## 3. The averaged model for binary suspension

Appendix A presents the averaging procedure which results in Eqs. (A16) and (A18). This derivation also allows for the calculation of the averaged suspension function, $f(c)$, given the suspension functions of each of the different types of particles. The derivation is presented in the most general form; allowing for continuous variation of properties within the injected particle population as determined by the vector, $r$. In this work we consider a two-particle system which corresponds to the case where all particles lie at two distinct values of $r$; i.e. the function $C(r,x,t)$ over $r$ is a sum of two delta functions.

For this system we can calculate the average suspended concentration from the suspended concentrations of the two different particles ($c_1=C(r_1,x,t)$, $c_2=C(r_2,x,t)$) from definition (A16):

$$c = c_1 + c_2 \tag{12}$$

For this work we consider the case where both particles exhibit traditional capture behaviour ($F(C(r))=\lambda_i C(r)$). As shown below, even for this case, the resulting averaged model exhibits a non-linear suspension function.

A key result of the *n*-particle model derivation is the discovery of a function $G(c_i)$ which is equal for all particles. This allows us to relate the suspended concentrations of the two particle species:

$$c_1 = c_1{}^0 \left( \frac{c_2}{c_2{}^0} \right)^{\frac{\lambda_1}{\lambda_2}} \tag{13}$$

Substituting the expression for $c_1$ into Eq. ( 12) results in:

$$c = c_1{}^0 \left( \frac{c_2}{c_2{}^0} \right)^{\frac{\lambda_1}{\lambda_2}} + c_2 \tag{14}$$

The parameters $c_1^0$ and $c_2^0$ are the dimensionless initial injected concentrations for the two particle populations such that $c_1^0 + c_2^0 = 1$. Throughout the injection, the ratio of the injected concentrations remains the same. Lastly, the suspension function is calculated from the two individual suspension functions using Eq. (A18):

$$f(c) = \lambda_1 c_1 + \lambda_2 c_2 \tag{15}$$

where the two individual concentrations are normalised by $c^0(0)$ and the two filtration coefficients are non-dimensionalised by multiplication by the system length, $L$, as per Eq. ( 7). Further in the text, references to both concentrations and filtration coefficients correspond to their dimensionless forms.

Using Eqs. ( 13) and ( 14) we can relate the two individual suspended concentrations to the averaged concentration, i.e. we can derive $c_1(c)$ and $c_2(c)$, which combined with Eq. ( 15) fully defines $f(c)$. Unfortunately, $c_1(c)$ and $c_2(c)$ cannot be derived as explicit expressions and so this procedure requires a numerical algorithm in order to solve the transcendental Eq. ( 15).

## 4. Behaviour of diluted suspensions

Let us briefly discuss the behaviour of the model at low suspended particle concentrations. In fact, this question can be interpreted as two different questions: either reducing the injected particle concentration (reducing $c^0$), or the successive reduction of the suspended concentration in the core through particle capture (reducing $c$). The process of reducing concentration by dilution prior to injection and reducing concentration through capture in the core are for this model starkly different. That is because the capture of particles in the core is selective; it removes the higher capture probability particles faster than lower probability particles. Thus, this kind of dilution shifts the population of suspended particles towards containing only type “2” particles. Fig. 2 shows it, where the ratio of type “1” and type “2” decrease from their injected ratio at $c=1$ to zero as $c$ tends to zero.

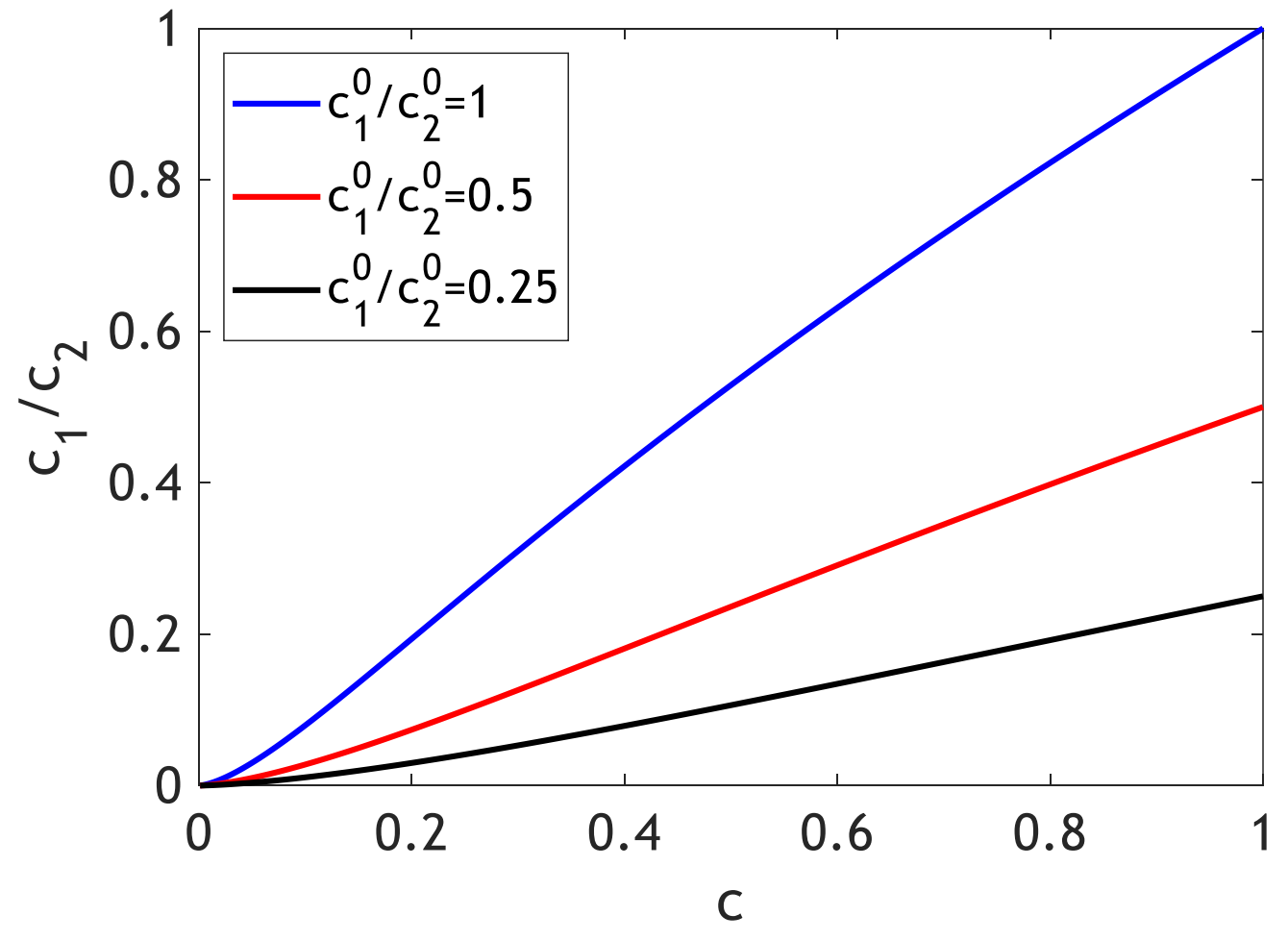

Fig. 2: Effect of diluting on a binary suspension. For this plot: $\Theta_\lambda=0.6$.

With a less diverse suspended particle population, the model starts exhibiting the traditional behaviour, i.e. *f(c)* becomes linear. This can be demonstrated by calculating the derivative of *f(c)* at low suspended concentrations (performed in Appendix B), which shows that in the limit of $c \to 0$, $\frac{\partial f(c)}{\partial c} = \lambda_2$.

Dilution of the injected sample however does not exhibit this effect. The derived dimensionless suspension function, *f(c)* does not contain the injected concentration $c^0(t)$ and thus its non-linearity is unaffected by the magnitude of the injected particle concentration. As a result, this model predicts a non-linear suspension function, and thus hyper-exponential retention profiles even for dilute injected suspensions, which has been reported in the literature [32].

## 5. Asymptotic expressions for suspension function

The inability to express *f(c)* explicitly can make calculations cumbersome. As a result, and in order to better understand the properties of the new model, we present in this section a number of asymptotic expressions for *f(c)* which allow for explicit representations. These expressions are derived by taking Taylor's series expansions for one of the constants in the model and assuming that it is small such that higher order terms can be neglected. An example derivation for the asymptotic expansion around low values of $c_1^0$ is presented in Appendix C. First- and second-order expansions involve neglecting terms higher than $\varepsilon$ and $\varepsilon^2$ respectively.

The first expansions involve the limiting cases where the binary mixture is comprised almost entirely of one type of particle. While the two particle populations are interchangeable, for consistency we maintain throughout the text that type "1" particles are captured faster than type "2" particles, such that $\lambda_1 > \lambda_2$. The first-order expansion for small parameter $c_1^0 = \varepsilon$ results in suspension function:

$$f(c) = \lambda_1(\varepsilon c^{\frac{\lambda_1}{\lambda_2}}) + \lambda_2(c - \varepsilon c^{\frac{\lambda_1}{\lambda_2}}) \quad (16)$$

This case corresponds to the limit in which the binary mixture contains only a very small fraction of type "1" particles. When the small parameter $\varepsilon$ is precisely zero, the filtration function becomes $f(c) = \lambda_2 c$ which corresponds to an injection of only type "2" particles.

Permitting higher powers of the small parameter $c_1^0 = \varepsilon$, we derive the second-order expansion:

$$f(c) = \lambda_1(\varepsilon c^{\frac{\lambda_1}{\lambda_2}} + \varepsilon^2 \frac{\lambda_1}{\lambda_2} c^{\frac{\lambda_1}{\lambda_2}}(1 - c^{\frac{\lambda_1}{\lambda_2}-1})) + \lambda_2(c - \varepsilon c^{\frac{\lambda_1}{\lambda_2}} - \varepsilon^2 \frac{\lambda_1}{\lambda_2} c^{\frac{\lambda_1}{\lambda_2}}(1 - c^{\frac{\lambda_1}{\lambda_2}-1})) \quad (17)$$

The first-order expansion for small concentrations of type "2" particles ($c_2^0 = \varepsilon$) results in the following suspension function:

$$f(c) = \lambda_1(c - \varepsilon c^{\frac{\lambda_2}{\lambda_1}}) + \lambda_2(\varepsilon c^{\frac{\lambda_2}{\lambda_1}}) \quad (18)$$

The corresponding second-order expansion gives:

$$f(c) = \lambda_1(c - \varepsilon c^{\frac{\lambda_2}{\lambda_1}} - \varepsilon^2 \frac{\lambda_2}{\lambda_1} c^{\frac{\lambda_2}{\lambda_1}}(1 - c^{\frac{\lambda_2}{\lambda_1}-1})) + \lambda_2(\varepsilon c^{\frac{\lambda_2}{\lambda_1}} + \varepsilon^2 \frac{\lambda_2}{\lambda_1} c^{\frac{\lambda_2}{\lambda_1}}(1 - c^{\frac{\lambda_2}{\lambda_1}-1})) \quad (19)$$

The asymptotic expansions for $c_1^0 = \varepsilon$ and $c_2^0 = \varepsilon$ are symmetric with respect to the filtration coefficients, $\lambda_i$.

Lastly, we have the case with an arbitrary mixture of the two particle populations and considering the limit in which the difference between their filtration coefficients is asymptotically small. We define a new dimensionless parameter which describes the difference between the capture properties of the two populations:

$$\theta_\lambda = \frac{\lambda_1 - \lambda_2}{\lambda_1}, \ \lambda_1 > \lambda_2 \quad (20)$$

This parameter varies between zero (equal capture rates) and one ($\lambda_1 >> \lambda_2$).

The first-order expansion for small parameter $\Theta_\lambda = \varepsilon$ results in the following suspension function:

$$f(c) = \lambda_1\left(c_1^0 c + \varepsilon c_1^0 c_2^0 c \ln(c)\right) + \lambda_2\left(c_2^0 c - \varepsilon c_1^0 c_2^0 c \ln(c)\right) \quad (21)$$

In the precise limit of $\Theta_\lambda=0$ we obtain a suspension function equal to $(\lambda_1 c_1^0 + \lambda_2 c_2^0)c = \lambda c$, meaning that we effectively have one population of particles which filters traditionally.

The second-order expansion for $\Theta_\lambda=\varepsilon$ results in the following suspension function:

$$f(c) = \lambda_1\left( c_1^0 c + \varepsilon c_1^0 c_2^0 c\ln(c) + \varepsilon^2 \left[ \ln(c)(c_1^0 c\ln(c) + c\ln(c)) + 2c_1^0 c\ln(c) + 2c\ln(c) + c_1^0 c\left(\ln(c)\right)^2 \right] \left( \frac{1}{c_1^0} + \frac{2}{c_2^0} \right)^{-1} \right) + \lambda_2\left( c_2^0 c - \varepsilon c_1^0 c_2^0 c\ln(c) - \varepsilon^2 \left[ \ln(c)(c_1^0 c\ln(c) + c\ln(c)) + 2c_1^0 c\ln(c) + 2c\ln(c) + c_1^0 c\left(\ln(c)\right)^2 \right] \left( \frac{1}{c_1^0} + \frac{2}{c_2^0} \right)^{-1} \right) \quad (22)$$

All six cases result in explicit expressions for the suspension function *f(c)*. Comparing the suspension functions derived from the asymptotic expansions to the original binary model by way of numerical calculations, Fig. 3a presents suspension functions calculated for different values of $c_1^0$ as well as the first- and second-order asymptotic expansions given in Eqs. ( 16) and ( 17). The graph shows that all suspension function curves are monotonically increasing with suspended concentration and for the binary model are all convex. Increasing $c_1^0$ increases the fraction of particles with higher capture rates and thus increases *f(c)* for all values of *c*.
The suspension functions calculated using asymptotic expansions show reasonable agreement with the binary model especially at low $c_1^0$, with a decrease in the alignment of the curves appearing for $c_1^0>0.4$. The region of validity of the asymptotic expansions varies depending on the values of $\lambda_1$ and $\lambda_2$. The asymptotic expansions show close agreement as the suspension concentration tends to zero. As discussed in Section 4, as $c\to 0$, the fraction of type '1' particles tends to zero, as they are selectively removed during capture due to their higher filtration coefficient. Thus, in the limit of small suspension concentrations, $c_1$ becomes small, which coincides with the asymptotic limit of the expansions shown here. The asymptotic expansions agree exactly at $c=1$. In fact, this is true for all asymptotic expansions presented in this work. The asymptotic expansions involve approximating the functions $c_1(c)$ and $c_2(c)$ (see Eq. (C11)). Thus, at any given value of *c*, the ratio $c_1/c_2$ in the asymptotic expansion model is not precisely equal to that of the original binary model. However, this ratio is fixed at $c=1$ at $\frac{c_1}{c_2} = \frac{c_1^0}{c_2^0}$ as determined by the boundary condition, and thus in any case the asymptotic expansions and binary model agree exactly for $f(c=1)$.
Fig. 3b shows the derivative of *f(c)* with respect to *c*. The solid curves exhibit the general property of the binary model that as *c* tends to zero, the derivative $\partial f(c)/\partial c$ tends to $\lambda_2$, as was discussed in Section 4. The asymptotic expansions for small $c_1^0$ honour this. The derivative of the binary model monotonically increases and is bounded within the interval:

$$\frac{\partial f(c)}{\partial c} \in \left[ \lambda_2, \frac{\lambda_1^2 c_1^0 + \lambda_2^2 c_2^0}{\lambda_1 c_1^0 + \lambda_2 c_2^0} \right] \quad (23)$$

The asymptotic expansions do not correctly describe the derivative at high values of *c*.

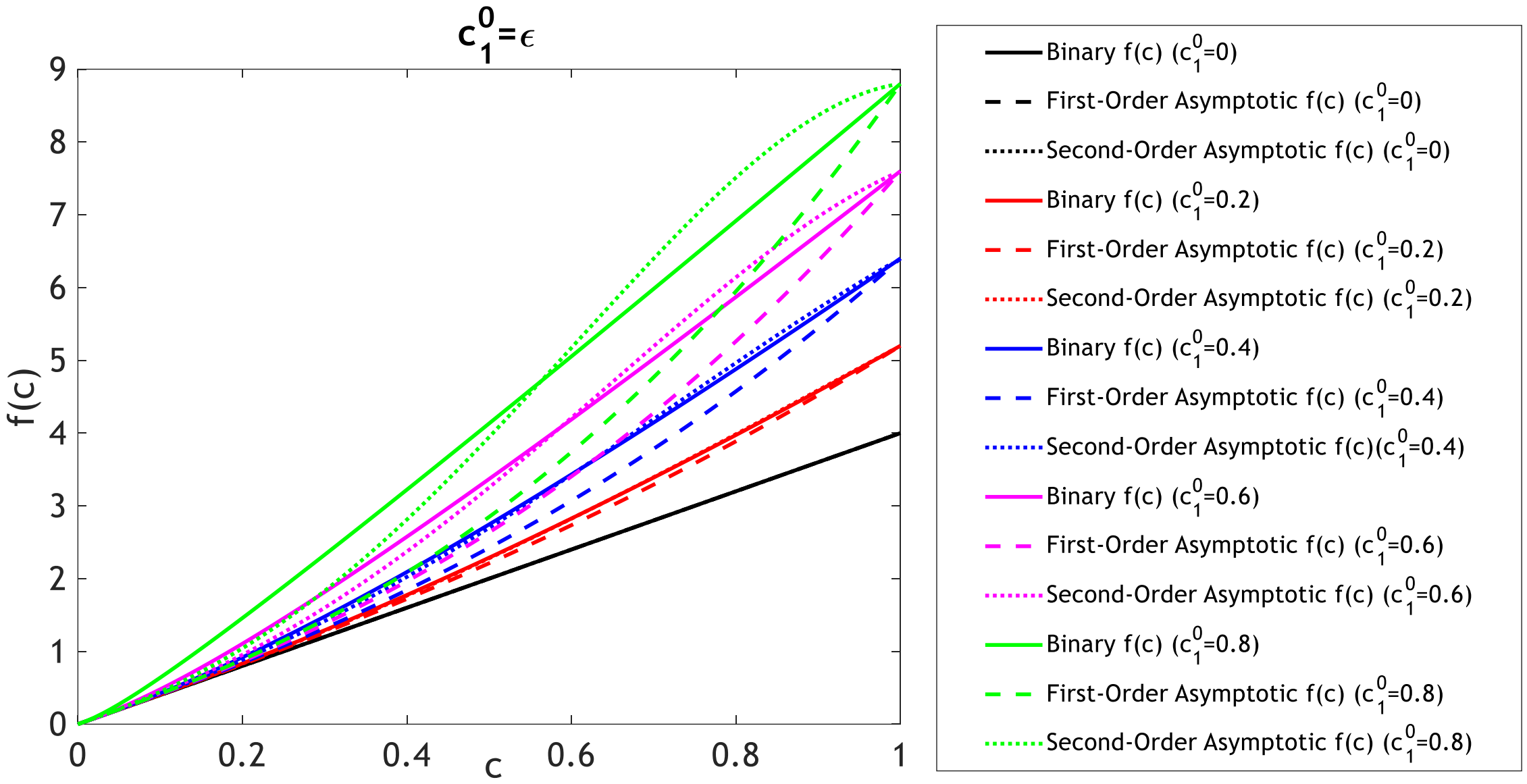

a)

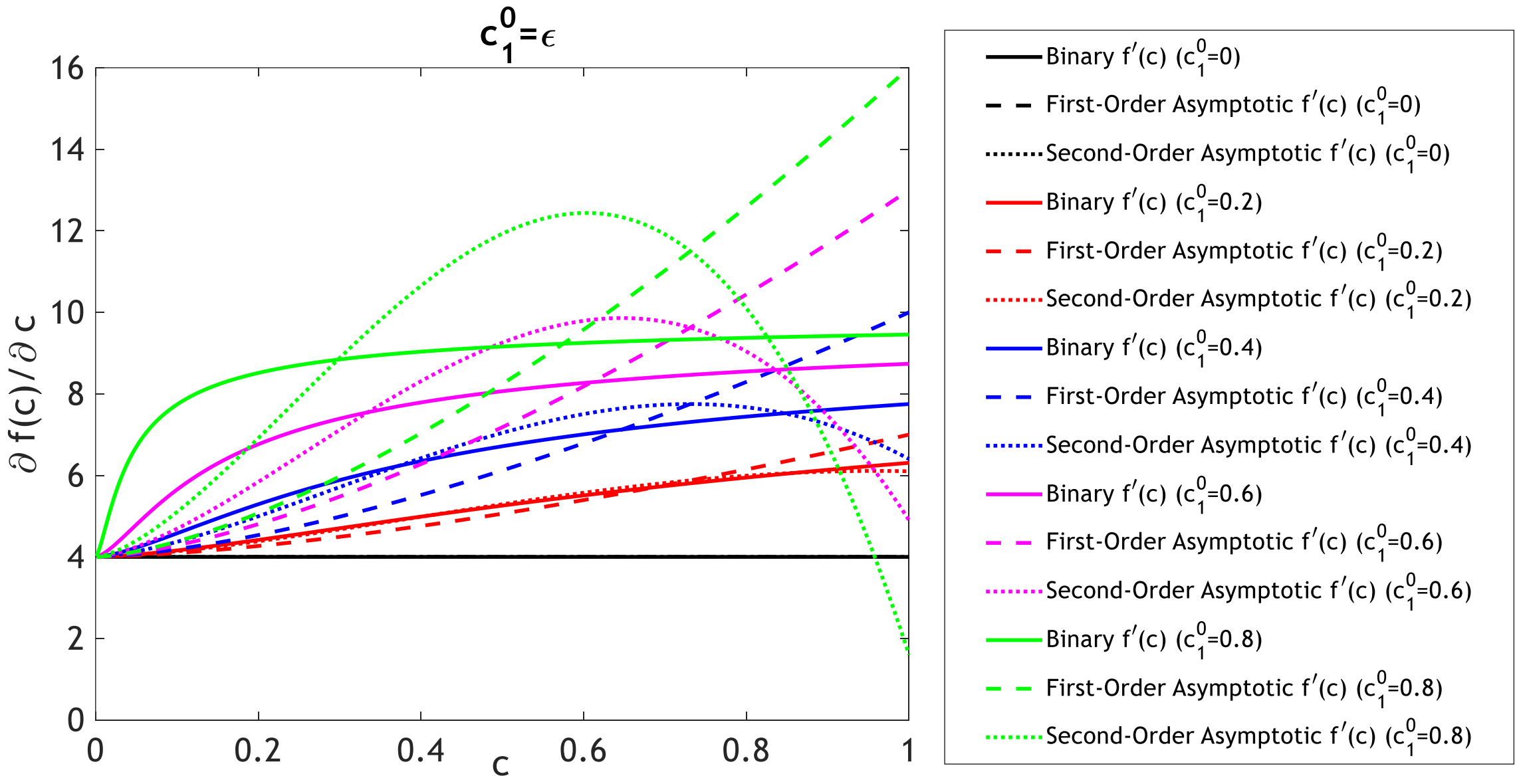

b)

Fig. 3: Sensitivity analysis on the suspension function when the initial fraction of the first population of colloids changes in a binary system. Dimensionless parameters for this plot are as follows: $\Theta_\lambda=0.6$, $\lambda_2=4$.

Fig. 4a presents several suspension functions with different values of $c_2^0$ alongside the two asymptotic expansions for low $c_2^0$ presented in Eqs. ( 18) and ( 19). Increasing $c_2^0$ decreases the value of $f(c)$ as it increases the proportion of particles with lower filtration coefficient.

The asymptotic expansions in the limit of small $c_2^0$ show good agreement at high values of $c$ with precise agreement at $c=1$ as explained above. The approximate curves begin to differ above $c_2^0=0.4$ and result in some cases in a negative value of $f(c)$ at low values of $c$ and in others in non-monotonic behaviour of the suspension function. These are to be considered errors induced by varying too far from the limit of the asymptotic expansions. The high deviation at low values of the suspension concentration follows from the property of the model that in this limit, the ratio of $c_1/c_2$ tends to zero. Due to the condition that $c_2^0$ be asymptotically small that is used to derive Eqs. ( 18) and ( 19), these expansions cannot adequately capture the behaviour that the concentration of type "2" particles greatly exceeds the concentration of type "1" particles at low $c$. The derivative presented in Fig. 4b shows the breakdown of the asymptotic expansions, as they fail to reproduce the lower limit of the derivative ($\lambda_2$) in contrast to the asymptotic expansions taken around small values of $c_1^0$.

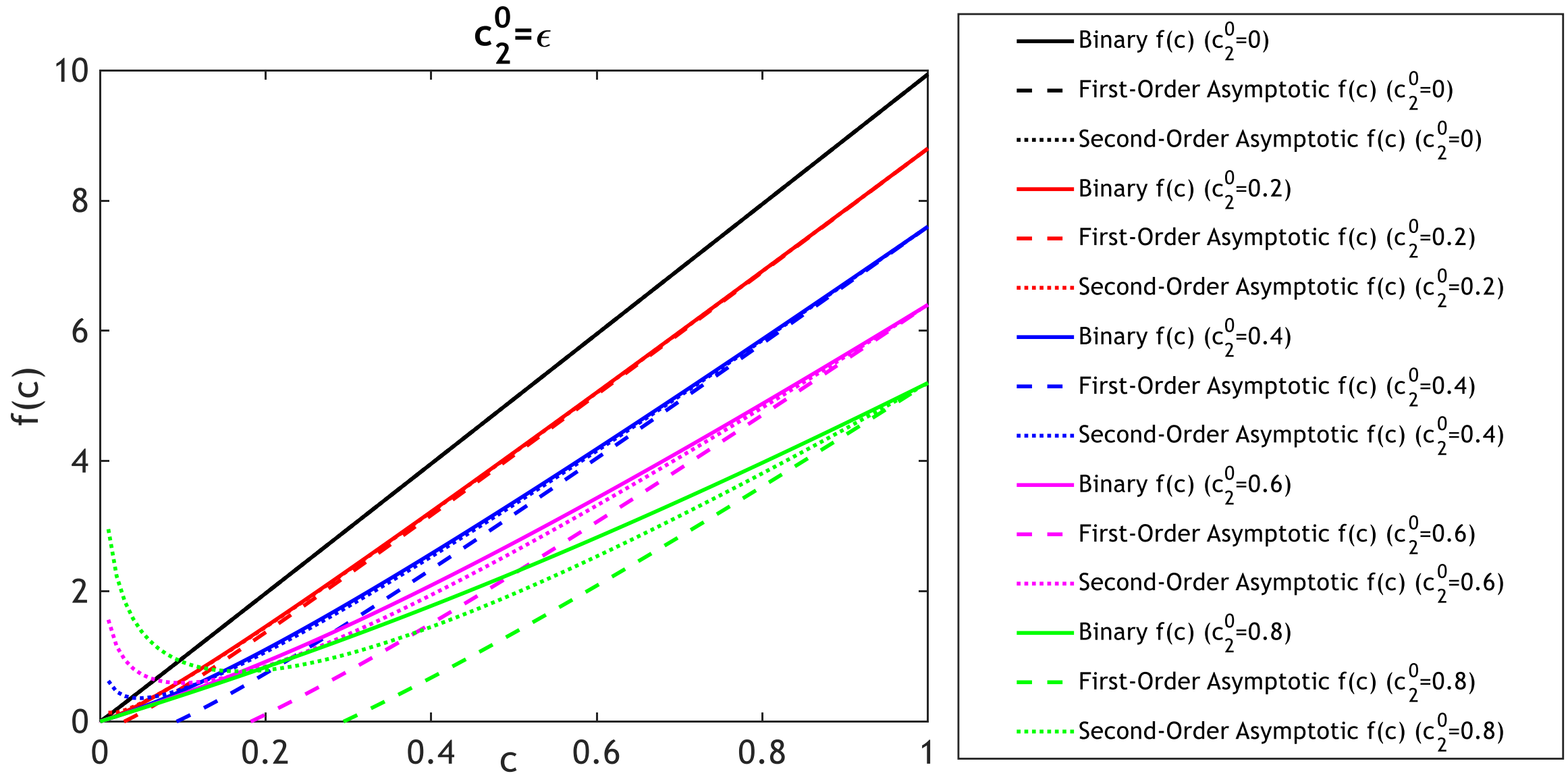

a)

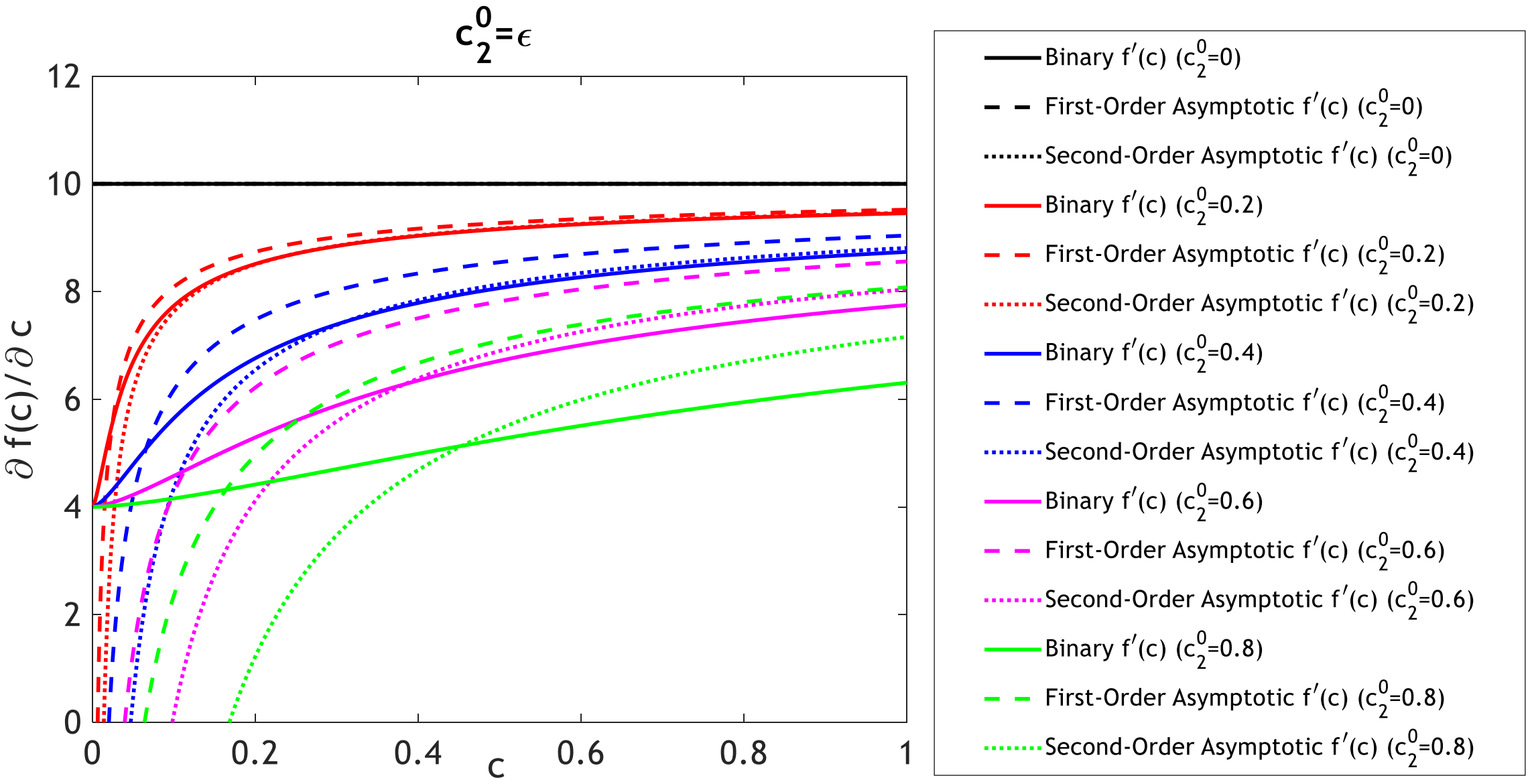

b)

Fig. 4: Sensitivity analysis on the suspension function when the initial fraction of the second population of colloids changes in a binary system. Dimensionless parameters for this plot are as follows: $\Theta_\lambda=0.6$, $\lambda_2=4$.

The final sensitivity analysis is performed to analyse the effect of the differences between the capture rates of the two particle species, as measured by $\Theta_\lambda$. For these calculations, $\lambda_1$ is fixed, and $\Theta_\lambda$ is changed by changing the value of $\lambda_2$. The decrease in $f(c)$ with increasing $\Theta_\lambda$ is a result of the choice to alter $\Theta_\lambda$ by decreasing $\lambda_2$. If instead $\Theta_\lambda$ was increased by increasing $\lambda_1$, then an increase in $\Theta_\lambda$ would increase $f(c)$.

The asymptotic expansions show high agreement, with deviations emerging above $\Theta_\lambda=0.4$. As with the first set of asymptotic expansions, those presented in Fig. 5a agree well at both $c=0$ and $c=1$.

Fig. 5b shows the derivative of the suspension function for different values of $\Theta_\lambda$. In this case, both the upper and lower bounds of the derivative change, as given explicitly by Eq. (B2). In the two extreme cases, $\Theta_\lambda=0$ and $\Theta_\lambda=1$ the upper bound of the derivative is equal to $\lambda_1$; initially this value decreases with $\Theta_\lambda$ and then at some point begins to increase, as observed in Fig. 5b as the derivative curve for $\Theta_\lambda=0.8$ exceeds the previous curve at high values of the suspended concentration.

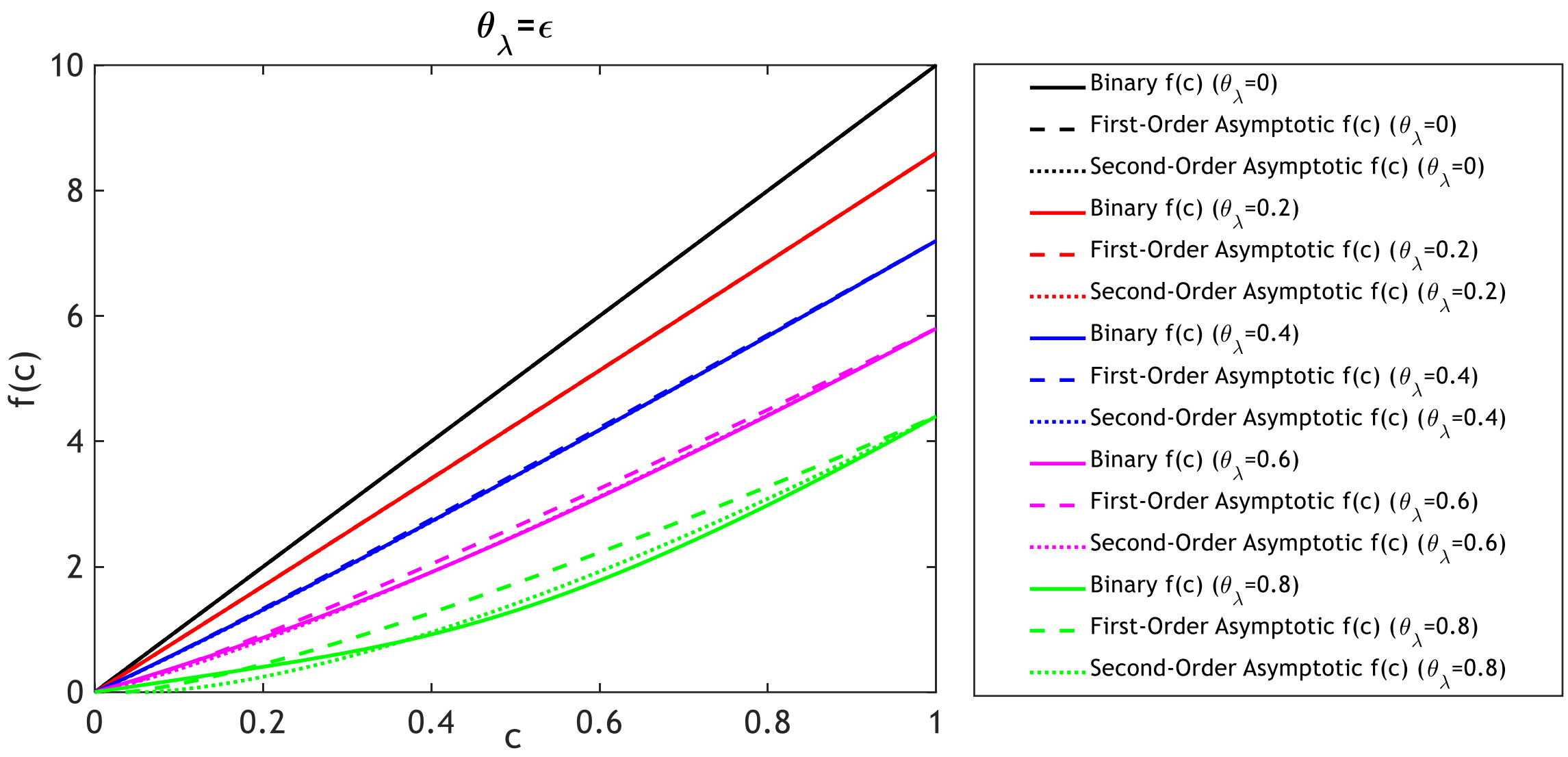

a)

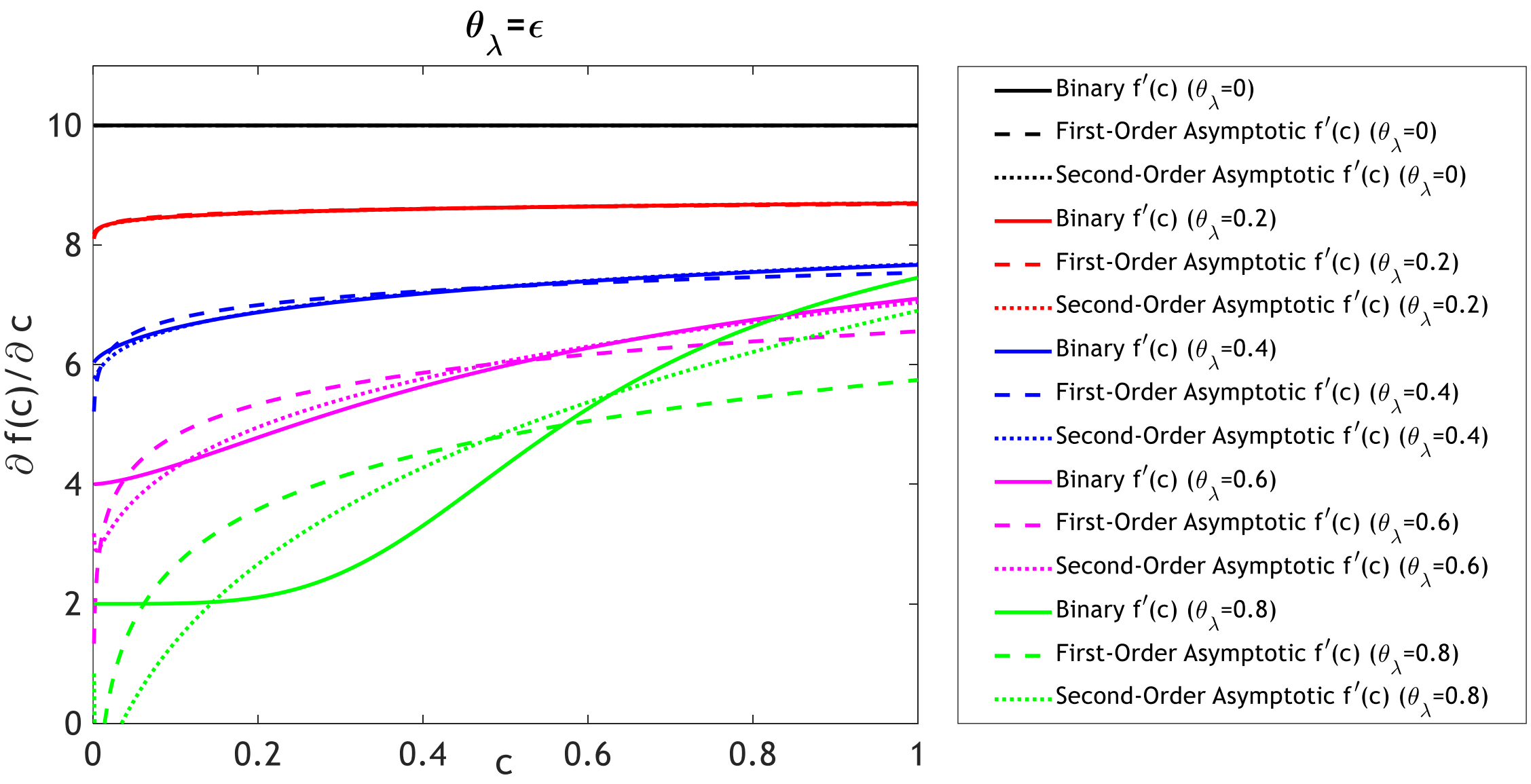

b)

Fig. 5: Sensitivity analysis on the suspension function when the dimensionless ratio between the filtration coefficients changes in a binary system. Dimensionless parameters for this plot are as follows: $c_1^0=0.3$, $\lambda_1=10$.

## 6. Exact solution for 1D upscaled colloidal transport

Eqs. ( 1) and ( 3)-( 5) subject to initial and boundary conditions ( 6) allow for an exact solution [41], [42]. The derivation is presented in Appendix D. For all $(x,t)$ values such that $x>t$, the core is unaffected by the injected particles and thus $c(x,t)=s(x,t)=0$. Along the line $x=t$ the suspended particle concentration experiences a shock, increasing from $c(x,t)=0$ ahead of this front, to some value $c^-(x,t=x)$ behind the front. Eq. (D9) in Appendix D presents the formula for calculating $c^-$ and Eq. (D17) presents the formula for calculating $c(x,t)$ for all $x<t$. The strained concentration is calculated from Eq. (D20) and the pressure drop and its normalised value (impedance, $J$) are calculated using Eqs. (D21) and (D22) respectively. Table 1 gives the solutions for $c(x,t)$ and $s(x,t)$.

Table 1: Exact solution for 1D particle transport for any arbitrary suspension function $f(c)$

| Flow zone | $c(x,t)$ | $s(x,t)$ |
|---|---|---|
| $x>t$ | 0 | 0 |
| $t>x>t-t_2$ | $\int_{c^-(x,x)}^{1}\frac{du}{f(u)}=x$ <br> $\int_{c_1^-(x,x)}^{c(x,t)}\frac{du}{(1-u)f(u)}=\frac{1}{s_m}(t-x)$ | $s(x,t)=s_m\left[\frac{c(x,t)-c_1^-(x,x)}{1-c_1^-(x,x)}\right]$ |
| $t-t_i>x>t-t_{i+1}, i>2$ | $\int_{c_{i-1}^-(x,x+t_{i-1})}^{c_i^-(x,x+t_i)}\frac{du}{(1-u)f(u)}=\frac{t-t_{i-1}}{s_m}$ <br> $\int_{c^-(x,x+t_i)}^{c(x,t)}\frac{du}{(1-u)f(u)}=\frac{t-t_i}{s_m}$ | $s\left(x,t\right)=s_m+\left[\frac{a_i-c\left(x,t\right)}{a_i-c_i^-\left(x,x+t_i\right)}\right]$ |

For a constant injected concentration at the inlet, we simply take the limit of the solution in Table 1 with $t_2\rightarrow\infty$ and thus only the solution in the first two rows is required.

For some particular forms of suspension function, we can derive explicit solutions for *c(x,t)* and *s(x,t)*. For example, for the traditional suspension function, where $f(c)=\lambda c$, if we inject a constant concentration, $c^0(t)=c^0$, the concentration of suspended particles just behind the front can be found by integrating Eq. (D9):

$$e^{-x\lambda}=c^-(x) \tag{24}$$

Then the suspended particle concentration can be calculated by integrating Eq. (D17):

$$c=\left(1-e^{-\lambda\frac{\tau}{s_m}}+e^{\left(-\lambda\frac{\tau}{s_m}+\lambda x\right)}\right)^{-1} \tag{25}$$

Finally the strained concentration can be calculated from Eq. ( 26) using the solution for $c^-(x,x)$ and *c(x,t)*:

$$s(x,t)=s_m\left(1-e^{-\lambda\frac{\tau}{s_m}}\right)\left(1-e^{-\lambda\frac{\tau}{s_m}}+e^{\left(-\lambda\frac{\tau}{s_m}+\lambda x\right)}\right)^{-1} \tag{26}$$

For some forms of suspension functions, concentrations need to be calculated numerically.

## 6.1 Behaviour of the solution

The behaviour of the analytical solution using suspension functions from the traditional model, binary model, and asymptotic model are presented in Fig. 6. The parameters used for this calculation are $\lambda_2=0.74$, $\lambda_1=22.08$, $c_1^0=0.3$, $\beta=10000$, $\phi=0.3$, $s_m=10$, and for the calculation of the impedance, *J*, we used $c^0=1000$ *ppm*.

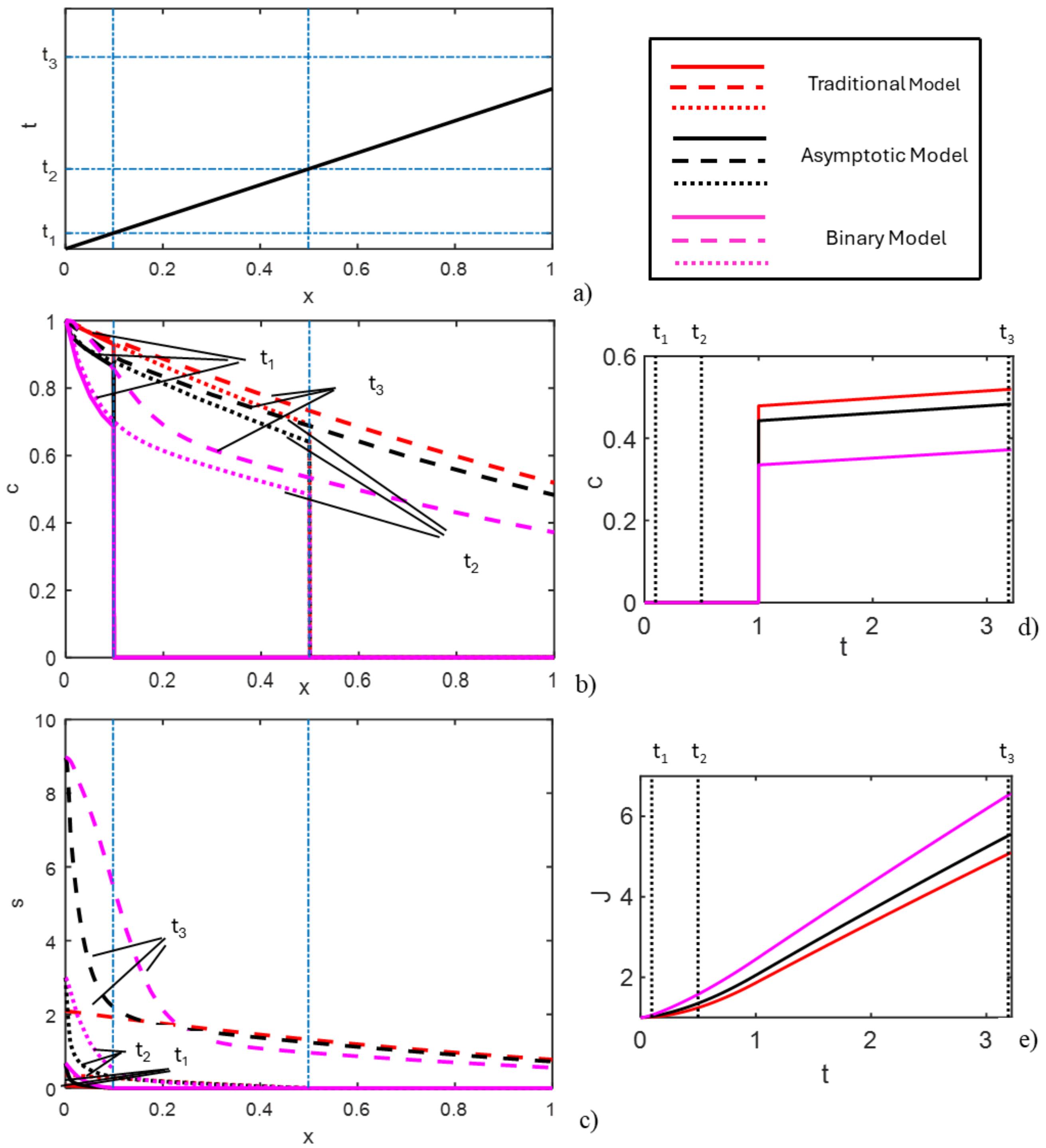

Fig. 6: Type curves from analytical modelling by the binary, asymptotic, and traditional models: (a) concentration front trajectory in (x,t)-plane; (b) suspension concentration profiles at three moments; (c) retention profiles; (d) breakthrough curves; (e) impedance curves.

The calculations are shown for three different times: two times, $t_1$ and $t_2$ before breakthrough, and one time, $t_3$ after it presented as solid, dotted, and dashed lines respectively. The *x-t* plane in Fig. 6a shows the x-coordinates of where the injected particle front lies at these three moments. The suspended particle concentration profiles presented in Fig. 6b show a monotonically decreasing function with *x*, which experiences a shock at the injected particle front down to $c=0$. The binary and asymptotic models exhibit higher curvature in $c(x)$ behind the front. In all cases the suspended concentration increases monotonically with time behind the injected particle front due to the Langmuir filtration function. The retained particle concentration curves are presented in Fig. 6c. For the parameters used in this calculation, both the binary and asymptotic models exhibit sharp declines in $s$ near the core inlet typical of hyper-exponential profiles reported in the literature. For all cases, the retained concentration increases monotonically with time and is zero ahead of the injected fines front. Fig. 6d and Fig. 6e show the breakthrough concentration and normalised pressure drop, $J$, two commonly measured variables during coreflooding tests. The breakthrough concentration is zero until $t=1$ and increases monotonically with time for all times $t>1$. The impedance increases monotonically with time from $t=0$, coinciding with the build-up of retained particles within the core. While not exhibited in these graphs due to the short time scale, as a result of the Langmuir filtration function, the suspended concentration tends to one, and the impedance stabilises at the point when $s(x)=s_m$ at all points in the core.

Figure S1 presented in the supplementary material repeat the sensitivity analysis presented in Section 5 for the log of the retained concentration profiles. Figure S1a for the approximation $c_1^0 = \varepsilon$ shows that for low $c_1^0$ the asymptotic expansions produce similar retained particle profiles, however at larger values of $c_1^0$, where the profiles begin to show significant hyper-exponentiality, the approximations begin to differ significantly from the binary model. The plots for $c_2^0 = \varepsilon$ (Fig. S1b) show a severe breakdown of the asymptotic expansions at high values of $c_2^0$, consistent with the $f(c)$ curves in Fig. 4a. Figure S1c showing the retained profiles for $\Theta_\lambda = \varepsilon$ shows close agreement between the binary model and the asymptotic expansions.

## 7. Validation of the exact solution

To validate the exact solution, it is compared with a numerical solution of the system of Eqs. ( 8) and ( 9) in Fig. 7. The numerical scheme used was Richtmyer's two-step variant of the Lax-Wendroff (LxW) method implemented in Matlab code provided by Shampine [43]. The numerical and analytical models are compared for three different ratios of the filtration coefficients, $\lambda_1$ and $\lambda_2$, and the two models agree almost exactly for both the retained concentration profiles (Fig. 7a) and the breakthrough concentration curves (Fig. 7b).

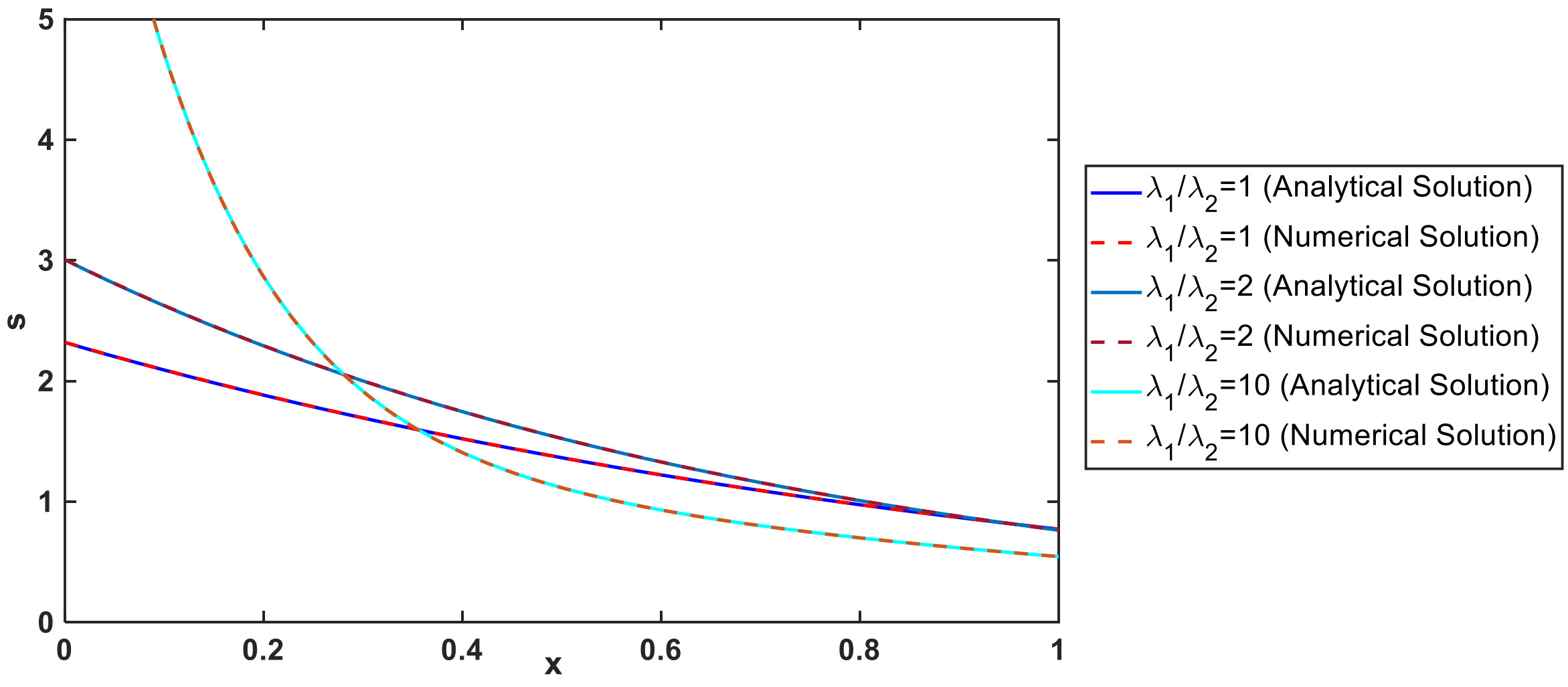

a)

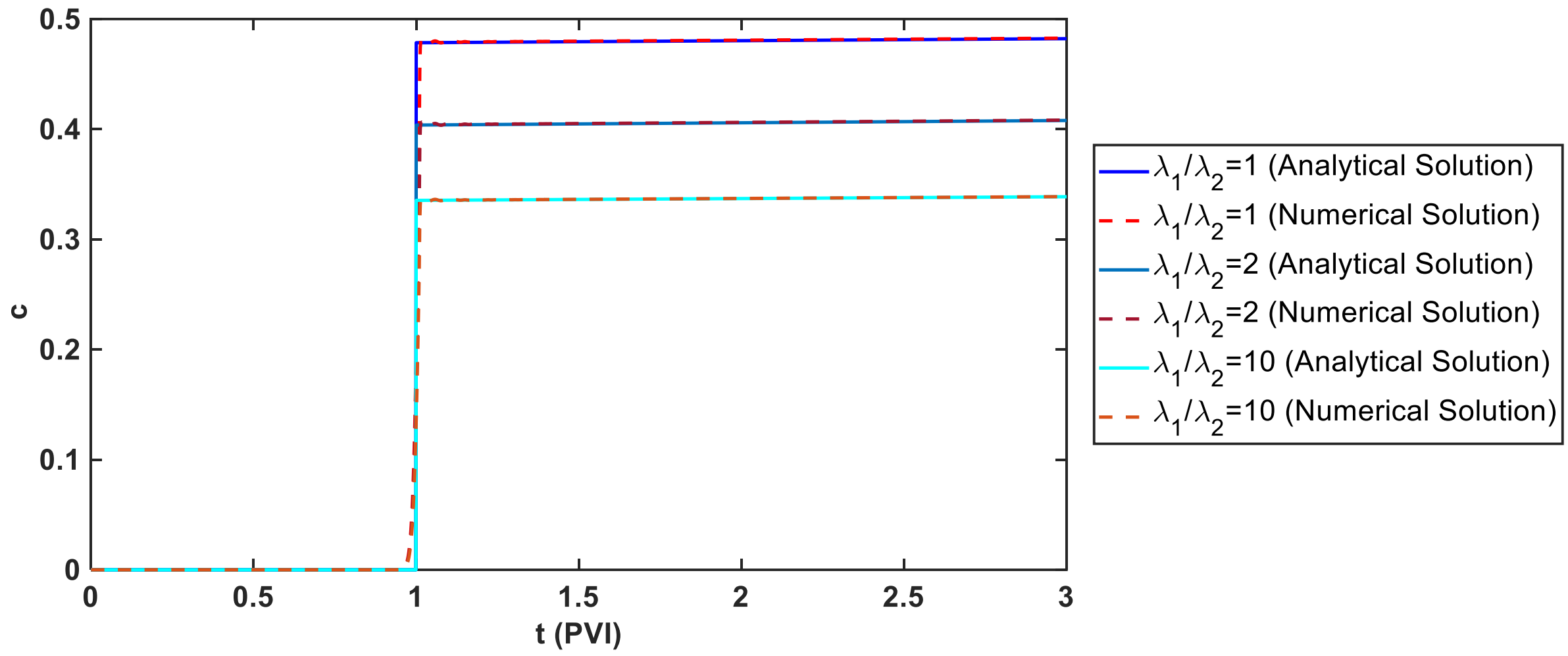

b)

Fig. 7: Validation of analytical solution with Shampine's code, sensitivity analysis on $\lambda_1/\lambda_2$. The calculations were performed for the following dimensionless parameters: $\lambda_2=0.74$, $c_1^0=0.3$, $s_m=100$.

## 8. Matching the laboratory coreflood data

In this section we match several sets of laboratory coreflood data with the traditional, binary, and asymptotic (Eq. ( 16)) models presented above. The experimental data matching by three models are performed using nonlinear least square method, which minimises the normalised deviation between the modelling and the experimental data. A gradient-based algorithm was implemented in Matlab (Mathworks, 2023) to perform the matching. In all cases, the data is tuned using the exact solutions (D9)-(D17) derived in Appendix D for a continuous injected concentration.

### 8.1 Matching the pressure drop

The first sets of laboratory data are taken from Ramachandran and Fogler's laboratory experiments [44]. In these experiments, 0.249 µm Polystyrene sulphate particles and membranes with 1 µm pores were used.

For these tests, only pressure drop data is available, which we transform into the normalised dimensionless pressure drop, or impedance, *J*. For the traditional model, the unknowns determined from matching are $(\beta, s_m, \lambda)$. For the binary and asymptotic models, while injected particle heterogeneity is assumed, the ratio of injected particles of each type is unknown. Thus, the injected concentration $c_1^0$ is treated as an unknown. The total list of unknowns for these two models is $(\beta, c_1^0, s_m, \lambda_1, \lambda_2)$. Three sets of experimental data are taken from this work, each with different injected particle concentrations, $c^0$. The only parameter that should physically differ between the tests is $c_1^0$. Thus, multiple tests are tuned simultaneously both to better improve the parameter estimation and more rigorously test the validity of the model. Two tests, $c^0 = 1235\ ppm$ and $c^0 = 720\ ppm$ are simultaneously tuned with $(\beta, s_m, \lambda)$ being matched for the asymptotic and binary models and $(\beta, c_1^0(1235ppm), c_1^0(720ppm), s_m, \lambda_1, \lambda_2)$ being tuned. The resulting tuned parameters, presented in Table 2, are then used to predict the outcome for the third data set $c^0 = 220\ ppm$. For the asymptotic and binary model, $c_1^0$ is tuned for this case. The results of the fitting are presented in Fig. 8, with fitting parameters and coefficient of determinations, $R^2$ for each test given in Table 2.

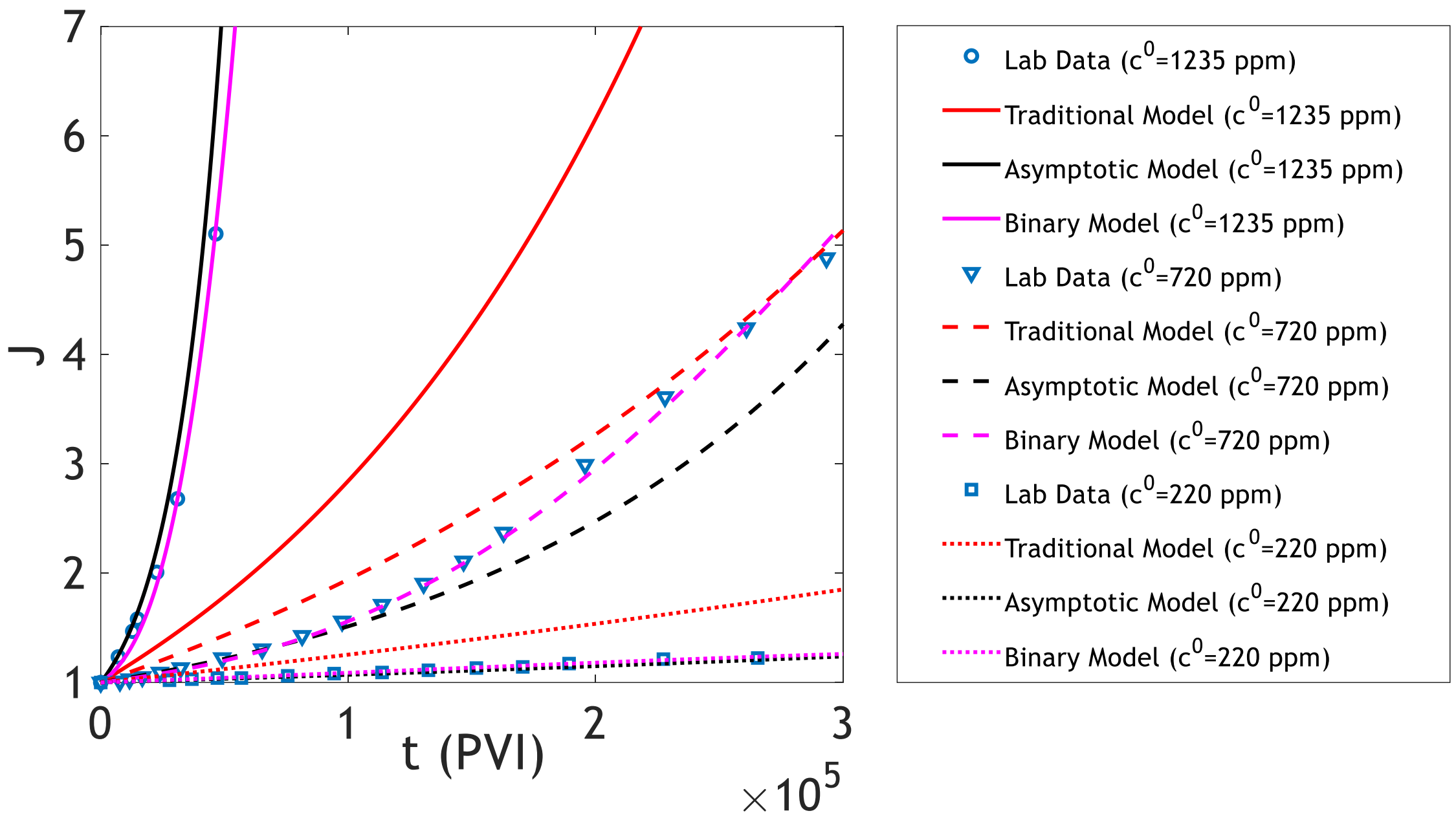

Fig. 8: Treatment of impedance lab data [44].

The tuning results show that the binary model exhibits the greatest agreement with the laboratory data. The asymptotic model shows reasonable agreement, but this begins to deviate at larger times. The traditional model shows a very poor fit, and the prediction at 220 ppm is very far from the laboratory data. Missing values of $R^2$ in Table 2 correspond to negative values, which we consider a failed match. The inability of the traditional model to match the data suggests that the capture behaviour is highly non-linear, likely exhibiting hyper-exponential retention profiles. Without measurements of *s(x)* this cannot be verified. The prediction at 220 ppm shows good agreement for both the binary and asymptotic model, indicating that the models align with the underlying physics phenomena which result in the increasing pressure drop across the core.

Table 2: Tuning parameters of Fig. 8.

| **Models** | **Test Conditions** | **β** | **$c_1^0$ [ppm]** | **$S_m$ [$L^{-3}$]** | **$\lambda_1$ [1/m]** | **$\lambda_2$ [1/m]** | **$R_p^2$** | **Dataset** |
|---|---|---|---|---|---|---|---|---|
| Traditional | $U = 4.8\times10^{-3}$ m/s, $c^0$= 1235 ppm | 5.02E+05 | | -4.61E-06 | 2.70E-01 | | - | Tuning |
| | $U = 4.8\times10^{-3}$ m/s, $c^0$= 720 ppm | | | | | | 0.957 | Tuning |
| | $U = 4.8\times10^{-3}$ m/s, $c^0$= 220 ppm | | | | | | - | Prediction |
| Asymptotic | $U = 4.8\times10^{-3}$ m/s, $c^0$= 1235 ppm | 4.77E+06 | 531.05 | -1.24E-07 | 1.11E-01 | 2.52E-04 | 0.882 | Tuning |
| | $U = 4.8\times10^{-3}$ m/s, $c^0$= 720 ppm | | 89.28 | | | | 0.962 | Tuning |
| | $U = 4.8\times10^{-3}$ m/s, $c^0$= 220 ppm | | 14.52 | | | | 0.914 | Prediction |
| Binary | $U = 4.8\times10^{-3}$ m/s, $c^0$= 1235 ppm | 1.07E+06 | 203.282 | -1.52E-07 | 4.55E-02 | 5.50E-03 | 0.995 | Tuning |
| | $U = 4.8\times10^{-3}$ m/s, $c^0$= 720 ppm | | 184.393 | | | | 0.998 | Tuning |
| | $U = 4.8\times10^{-3}$ m/s, $c^0$= 220 ppm | | 52.7511 | | | | 0.992 | Prediction |

An additional six *J(t)* curves have been tuned with the model. The resulting matches are presented in Figs. S2 and S3 and the matching parameters are given in Tables S1 and S2 in the supplementary material.

## 8.2 Matching suspension and retained concentrations

The next set of laboratory tests are taken from Yang et al.'s work [45]. Fig. 9 presents the results of colloidal injection into porous column packed by quartz sand with 425-600 μm. The porosity of the core was 0.47 and the column length and diameter in these tests were reported as 12 cm and 2.5 cm respectively.

The measured data in these tests included both breakthrough concentrations, and retention profiles measured after the tests completion. For each injection, both of these datasets are tuned simultaneously. The first dataset includes three tests performed at different values of pH. The tuning parameters for the traditional model are ($\beta$, $s_m$, $\lambda$) and for the binary and asymptotic models are ($\beta$, $c_1^0$, $s_m$, $\lambda_1$, $\lambda_2$). For these cases, each individual test is matched separately. The data and matched model curves for the breakthrough concentration, and the logarithm of the retention profiles are presented in Fig. 9a and Fig. 9b respectively.

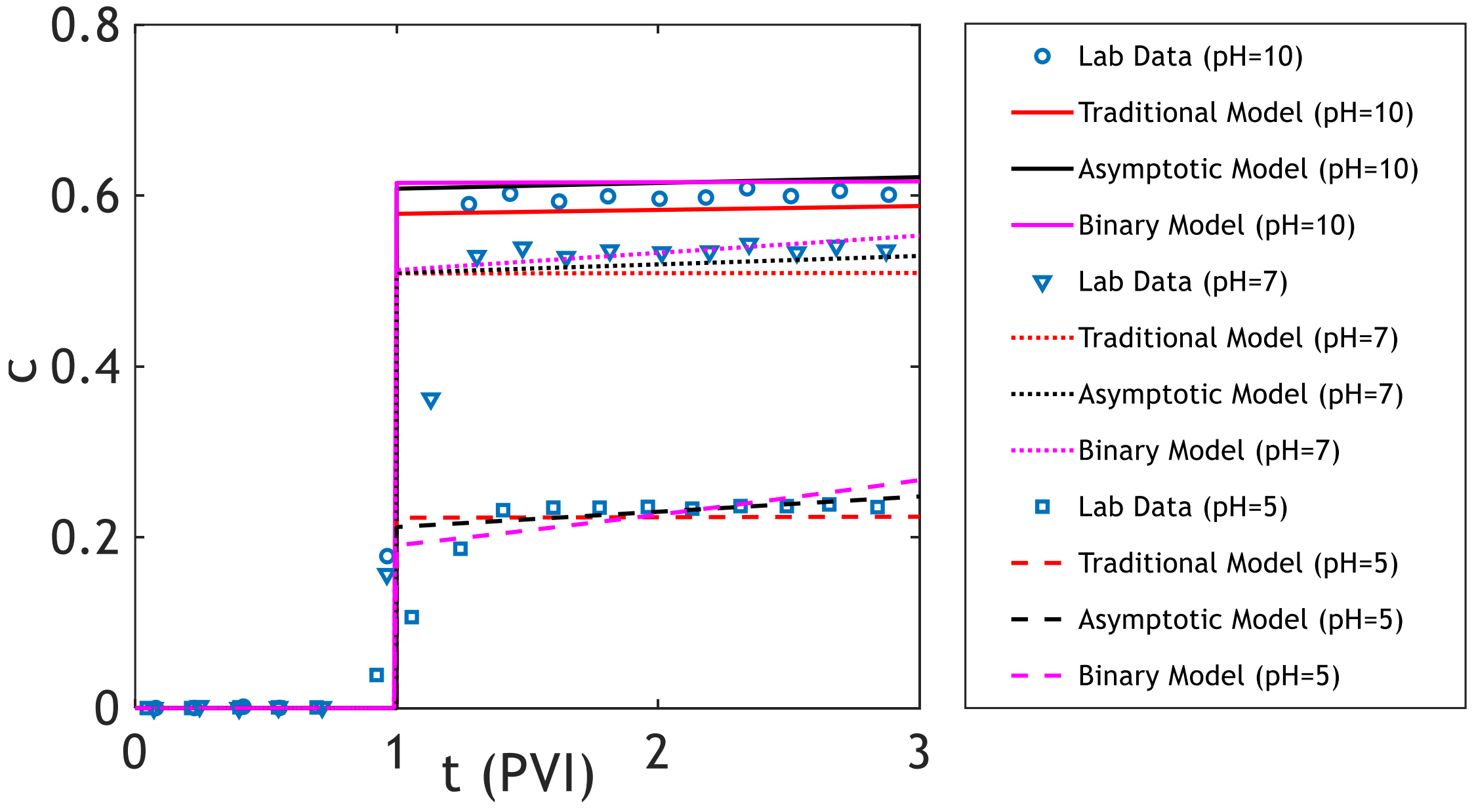

a)

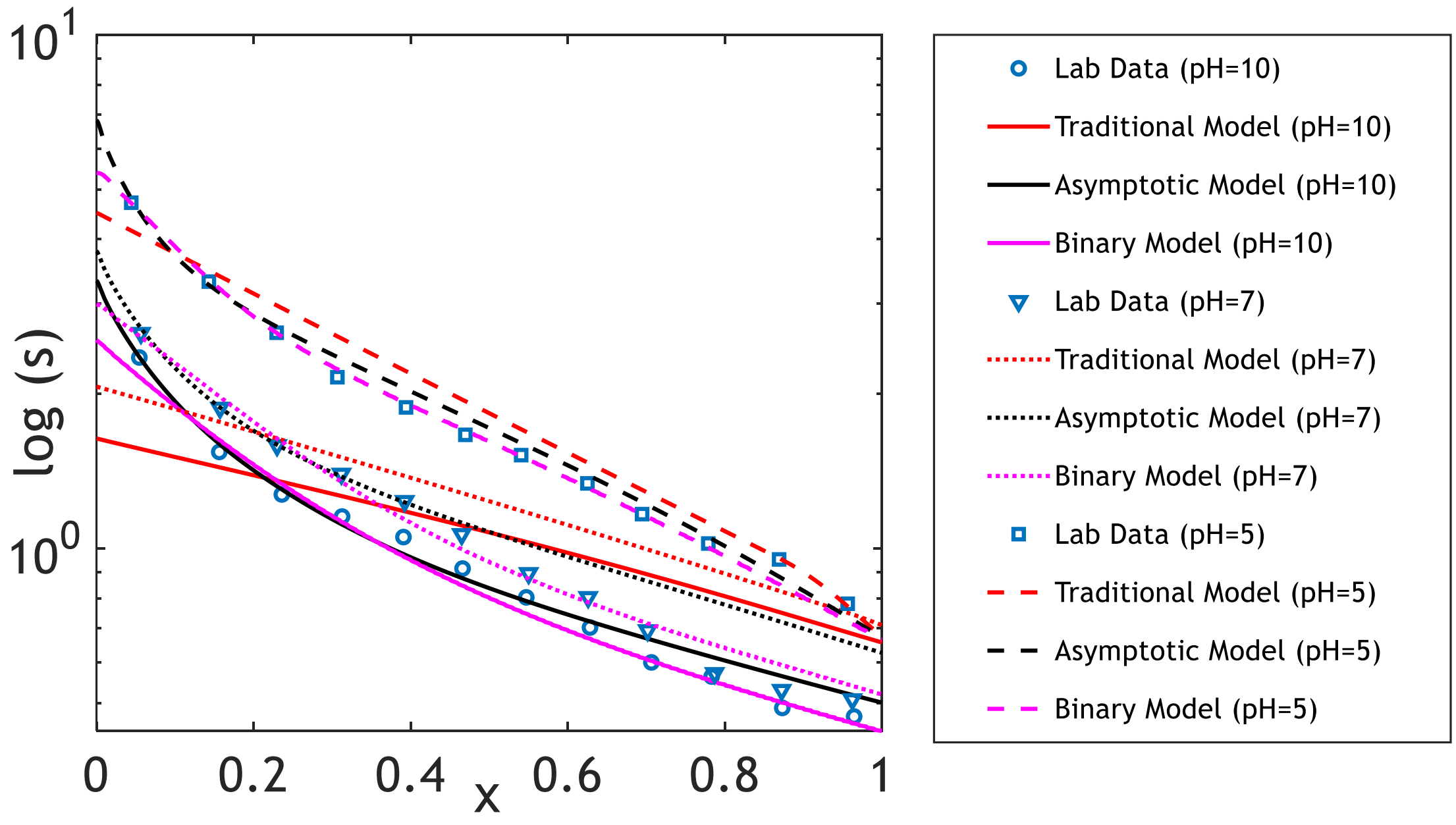

b)

Fig. 9: Simultaneous treatment of the suspended and retained concentrations of lab data [45].

The parameters obtained from fitting and the coefficients of determination are presented in Table 3. The breakthrough concentration curves in Fig. 9a show good agreement with the model for all three models in all three tests. The scattering of the data makes it difficult to discern whether the slightly increasing concentration in the binary model or the relatively constant concentration in the traditional model better represents the behaviour of the data. High coefficients of determination for the breakthrough concentration are observed for all models for all three tests. The retention profiles also show good agreement between the models; however, the traditional model notably fails to match the sharp decrease in *s(x)* near the core inlet. The *log(s)* vs *x* plot in Fig. 9b shows that near the core inlet the retention profiles for the datasets deviate slightly above the linear trend further in the core. This hyper-exponentiality is captured well by the binary and asymptotic models. The values of $R^2$ in Table 3 show that the binary model produces the highest quality of fit in all but one of the three injection cases. The filtration coefficients for both particles decrease monotonically with increasing pH which is in agreement with theoretical calculations of electrostatic forces which show a weakening of this attractive force with increasing pH [46].

Table 3: Tuning parameters of Fig. 9.

| Models | Test Conditions | $c_1^0$ [ppm] | $S_m$ [$L^{-3}$] | $\lambda_1$ [1/m] | $\lambda_2$ [1/m] | $R_c^2$ | $R_s^2$ |
|---|---|---|---|---|---|---|---|
| Traditional | $c^0$= 50 ppm | | 6.85E-04 | 4.563 | | 0.969 | 0.668 |
| Asymptotic | IS = 0.001 M | 9.63 | 3.29E-04 | 39.489 | 3.292 | 0.969 | 0.991 |
| Binary | BCs, pH=10 | 6.69 | 2.35E-03 | 34.996 | 2.88 | 0.969 | 0.979 |
| Traditional | $c^0$= 50 ppm | | 1.61E-02 | 5.630 | | 0.948 | 0.767 |
| Asymptotic | IS = 0.001 M | 6.245 | 3.54E-04 | 60.328 | 4.965 | 0.952 | 0.962 |
| Binary | BCs, pH=7 | 7.63 | 1.47E-04 | 50.803 | 4.189 | 0.952 | 0.99 |
| Traditional | $c^0$= 50 ppm | | 1.08E-02 | 12.642 | | 0.911 | 0.938 |
| Asymptotic | IS = 0.001 M | 4.89 | 3.46E-04 | 144.140 | 12.37 | 0.929 | 0.991 |
| Binary | BCs, pH=5 | 5.875 | 1.65E-04 | 146.879 | 12.772 | 0.939 | 0.998 |

The next dataset is taken from Li et al.'s paper [20]. In these tests, spherical soda lime glass beads have been used for packing the column with 20 cm length and 3.81 cm diameter. Porosity of the core was around 0.37. Spherical fluorescent carboxylate-modified polystyrene latex particles with diameter 1.1 μm were used.
Two different injections were done using different injected salinities. The data and matched curves are presented in Fig. 10, with tuned parameters given in Table 4.

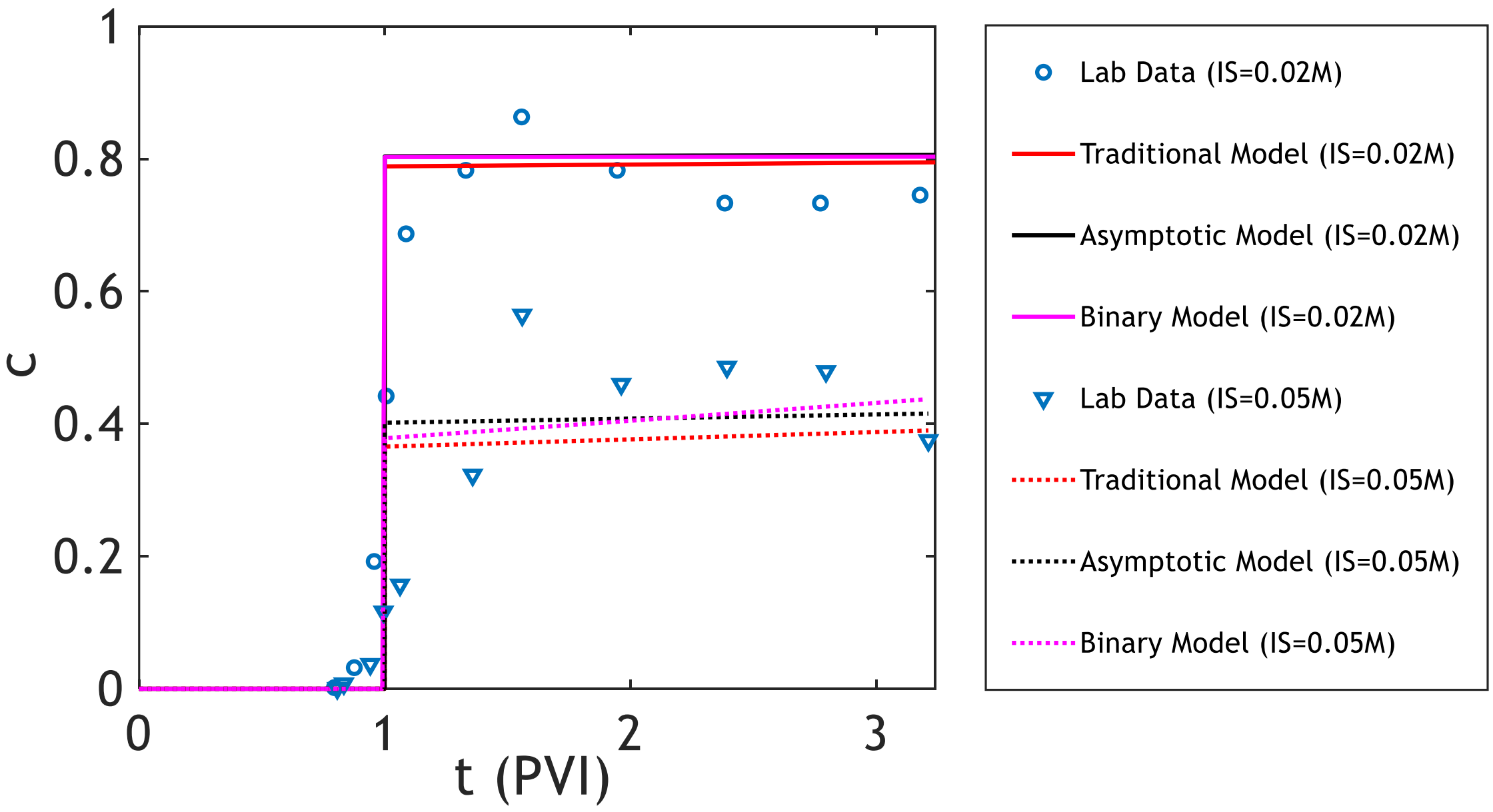

a)

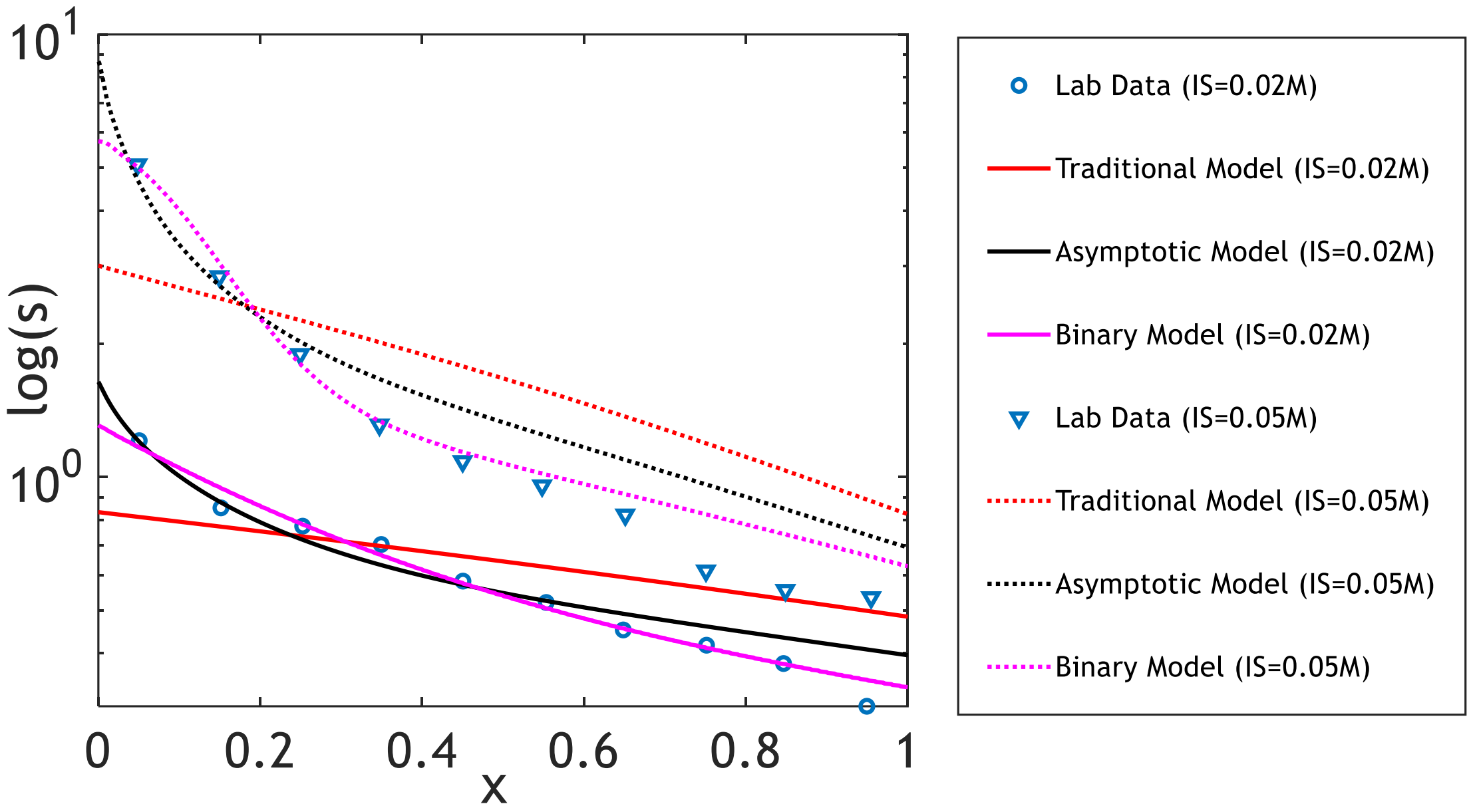

b)

Fig. 10: Simultaneous treatment of the suspended and retained concentrations of lab data [20].

As with the other tests, the suspended concentration is fit reasonably well by all models. The high scattering of the data results in lower values of $R^2$, however the models broadly follow the trend of the data. The retention profiles in Fig. 10b show that only the binary and asymptotic model adequately capture the behaviour of the model. The logarithm of the retention profile shows that the data exhibits highly hyper-exponential retention profiles, which are modelled well by the binary model. The coefficients of determination in Table 4 show that the asymptotic model also shows adequate agreement with the data. Increasing the salinity increases the attractive electrostatic force [40] which should increase the filtration coefficient. This is true for the binary and traditional model; however, the asymptotic model shows a slightly decreasing value of $\lambda_1$ when the salinity is increased from 0.02 M to 0.05 M.

Table 4: Tuning parameters of Fig. 10.

| **Models** | **Conditions of the Test** | $c_1^0$ [ppm] | $S_m$ [$L^{-3}$] | $\lambda_1$ [1/m] | $\lambda_2$ [1/m] | $R_c^2$ | $R_s^2$ |
|---|---|---|---|---|---|---|---|
| Traditional | $c^0$= 193 ppm | | 1.02E-03 | 1.917 | | 0.832 | 0.576 |
| Asymptotic | IS= 0.02 M | 36.415 | 2.30E-03 | 47.925 | 1.528 | 0.815 | 0.961 |
| Binary | | 28.986 | 7.20E-03 | 26.368 | 1.216 | 0.817 | 0.979 |
| Traditional | $c^0$= 193 ppm | | 1.55E-03 | 5.032 | | 0.727 | 0.585 |
| Asymptotic | IS= 0.05 M | 57.9 | 2.20E-03 | 44.175 | 3.885 | 0.735 | 0.955 |
| Binary | | 40.535 | 5.00E-04 | 84.647 | 3.680 | 0.775 | 0.991 |

An additional ten sets of breakthrough concentration and retention profiles have been matched and are presented in Figs. S4-S7 in the supplementary material. These curves also show good match and reasonable model parameters as seen in the Tables S3-S5. An additional curve is presented which shows the dependence of the filtration coefficient on the solution pH for the traditional, asymptotic, and binary models.

## 9. Discussions

The ability of the presented model to accurately match laboratory coreflood data exhibiting hyper-exponential retention profiles is confirmation that this phenomenon can be explained by heterogeneity in the injected particle population. The deficiency in existing models describing this process is that they are unable to describe the kinetics of this process where particles of all types contribute to 'depleting' the available capture sites; a phenomenon captured by the filtration function, *h(s)*. The advantages of this explanation for hyper-exponentiality as explored in this work are: first that this behaviour is exhibited even at low suspended concentrations, which has been reported in the literature [32], and secondly, that the non-linear behaviour arises from individual particle populations that filter traditionally, or linearly. Thus, the model does not require any more physical substantiation than those used for the traditional model of filtration in porous media.

One key assumption used in the derivation of the binary model (Appendix A) is that the two different particle populations occupy the same proportion of the available capture sites upon capture. As was shown in the litrerature [42], this assumption is not necessary, however in cases where different particles deplete the available capture sites at different rates, the upscaling of *2n* equations for each particle population results in not two equations as in this work, but three. This work has demonstrated that extending the model to this more general system is not necessary to capture hyper-exponential retention profiles. It has been shown that when using the general system, non-monotonic retention profiles can be produced [4], [47], which are another phenomenon that the traditional model for particle filtration in porous media cannot capture.

Another limitation of the model is the choice to use a binary particle model. In reality, the derivation in Appendix A allows for any general distribution of parameters. Adoption of a continuously heterogeneous particle population, as opposed to the two-particle system used in this work would allow for more general behaviour exhibited in the suspension function, *f(c)*. In addition, the choice of a Langmuir filtration function, *h(s)* is convenient due to the availability of an exact solution, however more general forms could be used including those which predict asymptotic stabilisation of particle capture [48], as opposed to the finite stabilisation exhibited by the model given in Table 1.

The application of the model in this work involved the inverse modelling of laboratory coreflood effluent suspended concentration and pressure drop data in order to determine the parameters of the binary model. In some applications it is important to predict the filtration properties of mixtures of different colloids whose individual filtration coefficients are known. As the model presented in this work explicitly describes the influence of the filtration coefficients of each particle, it can be used to make these predictions. Such a process could be used to optimize the injected particle composition in order to maximise or minimise the hyper-exponentiality of retention profiles.

The micro scale model (A1-A4) accounts for probabilistic distributions of multiple particle parameters, like sizes, aspect ratios, surface roughness, zeta-potentials, etc., numbered in Eq. (A1) as 1,2…n. Yet, stochastically distributed microscale porous-media parameters – pore size and form distributions, zeta-potentials, and non-uniform distributions of those properties over the surface – are not accounted for. Stochastic rock parameters can be captured by random walk models [17], Boltzmann's equation [38], continuous upscaling [33], [49], percolation and effective-medium theories [39], and numerical 3D modelling [50]. Fines transport by interphase surfaces and its capture by the rock can be accounted for using equations for two-phase flows with separating surfaces [51], [52]. For the above upscaled models, in large-scale approximation, the dissipative and non-equilibrium effects are negligible if compared with the advective mass fluxes [53]–[55]. This yields hyperbolic systems of conservation laws, where the exact solutions can be found by nonlinear modifications of Riemann invariants and potentials [41], [53], [56], like in Appendix D. The dissipative and non-equilibrium effects can be incorporated into large-scale solutions by matching the asymptotic expansions [54], [55].

## 10. Conclusions

Exact averaging (upscaling) of multicomponent colloidal flow in porous media allows concluding as follows. In the upscaled model for multicomponent colloidal transport with total suspended concentration *c*, where different retained particles occupy the same number of vacancies, nonlinear suspension function *f(c)* in the expression for capture rate substitutes the multiplier *c* in the traditional model. For example, for binary mixture, where each population filters by the traditional model, the upscaled single-population colloid filters with non-linear suspension function *f(c)*.

In the limit of vanishing suspended concentration, the suspension function tends to $f(c) = \lambda_2 c$. In the progression of the system from $c = 1$ to $c = 0$, particle capture preferentially removes the higher capture particles, which in the limit of $c \to 0$, results in a particle population entirely composed of the lower capture particles, resulting in traditional capture behaviour.

The non-linearity of the suspension function is independent of the total injected concentration, $c^0$, meaning that non-linear capture and hyper-exponential retention profiles are exhibited by the binary model even for very dilute suspensions.

The suspension function in the binary model cannot be expressed explicitly as a function of the averaged suspension concentration. As such, asymptotic expansions in an explicit form are derived for three limiting cases of the model: low large-particle concentration ($c_1^0 \to 0$), low small-particle concentration ($c_2^0 \to 0$), and small difference between the filtration coefficient of two populations ($\theta_\lambda \to 0$). Second-order expansions show better agreement with the binary model than the first-order expansions.

The 1-D flow model for colloidal suspension flow in porous media with any arbitrary non-linear suspension function and injected concentration versus time, but linear filtration function, allows for an exact solution. It contains the explicit expressions for the suspended and retained concentrations, and the pressure drop.

The model exhibits high agreement with simultaneous matching of suspended concentration histories and retained concentration profiles. The match with the tests measuring pressure drops across the cores is also high.

The traditional model cannot capture high degree of curvature of the pressure drop across the core $J(t)$, while the asymptotic and binary models can.

The traditional model fails to produce high agreement during simultaneous fitting of outlet concentration and retention profiles, especially in cases where the retention profiles are hyper-exponential. The asymptotic model can capture this behaviour and shows a high quality of fit for all cases. The binary model shows the highest level of agreement.

The obtained tuned parameters resulting from matching impedance and BTC and HERP curves simultaneously are within the common ranges.

## Appendix A. Derivation of upscaled model for multicomponent suspension flow

This Appendix derives an extended model by upscaling the micro-scale (deep bed filtration) DBF equations, where the particles are distributed by any property like radius, shape factor, z-potential, density, surface charge, etc. Let $r$ be a vector of particle properties $r=(r_1,r_2...r_n)$ and use the following notation for integral over all $r_i$:

$$\int_0^\infty u(r,x,t)\,dr = \int_0^\infty\int_0^\infty...\int_0^\infty u(r_1,r_2...r_n,x,t)\,dr_1 dr_2...dr_n \tag{A1}$$

Average suspension and retention concentrations are:

$$c(x,t)=\int_0^\infty C(r,x,t)\,dr,\ \ s(x,t)=\int_0^\infty S(r,x,t)\,dr \tag{A2}$$

Mass balance for suspended and retained particles is valid for each population with given property $r$.

$$\frac{\partial\left(\phi C(r,x,t)+S(r,x,t)\right)}{\partial t}+U\frac{\partial C(r,x,t)}{\partial x}=0 \tag{A3}$$

We consider filtering of different particle populations in the same rock, so the filtration function is the same for all populations, but suspension function depends on the particle radius [37], [57]:

$$\frac{\partial S(r,x,t)}{\partial t}=h\left(\int_0^\infty S(r,x,t)\,dr\right)F\left(C(r,x,t),r\right)U \tag{A4}$$

Here the functional *F(C,r)* accounts for interaction between all populations that corresponds to different particle properties $r=(r_1, r_2...r_n)$ [10], [13], [49].

Then we define the following dimensionless parameters.

$$c^0=\int_0^\infty C^0(r)\,dr,\ C\to\frac{C}{c^0},\ S\to\frac{S}{\phi c^0},\ F(C,r)\to\frac{F(C,r)L}{c^0},$$

$$C^0(r)\to\frac{C^0(r)}{c^0},\ r\to\frac{rc^0}{\int_0^\infty rC^0(r)\,dr} \qquad \text{(A5)}$$

Equations (A3) and (A4) takes the form:

$$\frac{\partial\left(C(r,x,t)+S(r,x,t)\right)}{\partial t}+\frac{\partial C(r,x,t)}{\partial x}=0 \qquad \text{(A6)}$$

$$\frac{\partial S(r,x,t)}{\partial t}=h(s)F\left(C(r,x,t),r\right) \qquad \text{(A7)}$$

System of two Eqs. (A6) and (A7) subject to initial and boundary conditions that correspond to colloid injection into clean porous bed defines two unknowns *C(r,x,t)* and *S(r,x,t)*. The micro-scale system (A6) and (A7) contains the model phenomenological functions of filtration and suspension.

$$h=h(s),\ F=F(C,r),\ r=(r_1,r_2...r_n) \qquad \text{(A8)}$$

Using timely variable $\tau$ and the corresponding mapping in system (A6) and (A7) yields:

$$\frac{\partial S(r,x,\tau)}{\partial\tau}+\frac{\partial C(r,x,\tau)}{\partial x}=0$$

(A9)

$$\frac{\partial S(r,x,\tau)}{\partial\tau}=h(s)F\left(C(r,x,\tau),r\right) \qquad \text{(A10)}$$

Substituting Eq. (A10) in Eq. (A9) yields:

$$\frac{\partial C(r,x,\tau)}{\partial x}=-h(s)F\left(C(r,x,\tau),r\right) \qquad \text{(A11)}$$

As it follows from clean-bed initial conditions and characteristic form of Eq. (A11), $C=S=0$ for $\tau<0$.

Separating variables in Eq. (A11) at $\tau>0$ yields:

$$\frac{\partial G\left(C(r),r\right)}{\partial x}=-h\left(\int_0^\infty S(r,x,\tau)\,dr\right),\ G\left(C(r),r\right)=\int_{C^0(r)}^{C(r)}\frac{du}{F(u,r)} \qquad \text{(A12)}$$

Right hand side of Eq. (A12) is independent of *r*, so its value for any arbitrary *r* is equal to that at the mean radius $r=(1,1...1)$:

$$\frac{\partial G\left(C(r),r\right)}{\partial x}=\frac{\partial G\left(C(1),1\right)}{\partial x} \qquad \text{(A13)}$$

Integrate both sides of Eq. (A13) in *x* characteristic line $\tau=\tau_0$. Functions *G(C(r), r)* are equal zero for any $r>0$ at $x=0$, so the integration constant is also zero. Therefore, along the characteristic line $\tau=\tau_0$, the following equality takes place:

$$\tau=\tau_0:\ G\left(C(r),r\right)=G\left(C(1),1\right) \qquad \text{(A14)}$$

By using the inverse function in Eq. (A14), express concentration *C(r,x,t)* via concentration *C(1,x,t)*:

$$C(r)=g\left(C(1),r\right),\ g=G^{-1}(C,r) \qquad \text{(A15)}$$

Now averaged concentration $c(x,\tau)$ can be expressed via the concentration of average-size particle $C(1,x,\tau)$, and vice versa, using the inverse function:

$$c=\int_0^\infty g\left(C(1),r\right)dr,\; C(1)=a(c) \qquad \text{(A16)}$$

Integrate both sides of Eq. (A11) in *r* and take *r*-independent function *h* out of the integral.

$$\frac{\partial c(x,\tau)}{\partial x}=-h\left(s(x,\tau)\right)\int_0^\infty F\left(C(r,x,\tau),r\right)dr \qquad \text{(A17)}$$

Integral in right hand side of Eq. (A17) accounting for expression (A15) becomes:

$$\int_0^\infty F\left(C(r,x,\tau),r\right)dr=\int_0^\infty F\left(g\left(C(1,x,\tau),r\right),r\right)dr = e\left(C(1,x,\tau)\right)=e\left(a\left(c(x,\tau)\right)\right)=f(c) \qquad \text{(A18)}$$

Eq. (A10) accounting for averaged relationship (A18) becomes:

$$\frac{\partial s(x,\tau)}{\partial \tau}=h\left(s(x,\tau)\right)f(c) \qquad \text{(A19)}$$

The averaged suspension function *f(c)* is the result of upscaling of micro-scale suspension function *F(C,r)* that is individual for each population.

Integration of both parts of Eq. (A9) in *r* yields

$$\frac{\partial s}{\partial \tau}+\frac{\partial c}{\partial x}=0 \qquad \text{(A20)}$$

The derived macro-scale system (A19) and (A20) contains the model phenomenological functions of filtration *h(s)* and of suspension *f(c)*. The upscaled suspension function depends not only on the suspension functions for each population, but also on the injected concentrations $C^0(r)$. So, the modified model can be obtained by averaging of micro-scale system with multiple distributed particle properties.

## Appendix B. Calculations of first and second derivatives of suspension function

Consider binary colloid described by the traditional DBF model with constant filtration coefficients $\lambda_1$ and $\lambda_2$, and $c_1^0$ as the injected concentration of the first component in the overall colloid.

From applying chain rule on Eq. ( 15) we can calculate the first derivate of *f(c)* with respect to *c*:

$$\frac{\partial f(c)}{\partial c}=\frac{\partial f(c)}{\partial c_1}\frac{\partial c_1}{\partial c}+\frac{\partial f(c)}{\partial c_2}\frac{\partial c_2}{\partial c} \qquad \text{(B1)}$$

The terms $\partial f(c)/\partial c_1$ and $\partial f(c)/\partial c_2$ calculated from Eq. ( 15) are equal to $\lambda_1$ and $\lambda_2$ respectively. The derivative of the concentration of type "2" with respect to the average concentration is determined by taking the derivative of Eq. ( 14) with respect to *c* and rearranging for $\partial c_2/\partial c$. $\partial c_1/\partial c$ can be calculated following the same procedure with changing Eq. ( 14) to become a function of $c_1$ instead of $c_2$.

Substituting the final expressions into Eq. (B1), we obtain:

$$\frac{\partial f(c)}{\partial c}=\lambda_1\left(1+\frac{\lambda_2}{\lambda_1}\frac{c_2^{\,0}}{\left(c_1^{\,0}\right)^{\frac{\lambda_2}{\lambda_1}}}\left(c_1\right)^{\frac{\lambda_2}{\lambda_1}-1}\right)^{-1}+\lambda_2\left(1+\frac{\lambda_1}{\lambda_2}\frac{c_1^{\,0}}{\left(c_2^{\,0}\right)^{\frac{\lambda_1}{\lambda_2}}}\left(c_2\right)^{\frac{\lambda_1}{\lambda_2}-1}\right)^{-1} \qquad \text{(B2)}$$

When, $c\rightarrow 0$, using the fact that $\lambda_2<\lambda_1$, Eq. (B2) becomes:

$$\frac{\partial f(0)}{\partial c}=\lambda_2 \qquad \text{(B3)}$$

Again, applying the chain rule, we can derive the second derivative of *f(c)* with respect to *c* in a binary system:

$$\frac{\partial^2 f(c)}{\partial c^2} = \frac{\partial^2 f(c)}{\partial c_1^{\,2}}\left(\frac{\partial c_1}{\partial c}\right)^2 + \frac{\partial f(c)}{\partial c_1}\frac{\partial^2 c_1}{\partial c^2} + \frac{\partial^2 f(c)}{\partial c_2^{\,2}}\left(\frac{\partial c_2}{\partial c}\right)^2 + \frac{\partial f(c)}{\partial c_2}\frac{\partial^2 c_2}{\partial c^2} \tag{B4}$$

Given that the first derivatives of *f(c)* with respect to each concentration are constants, the second derivatives are zero. The second derivatives of each individual concentration with respect to the average concentration are calculated from the first derivatives using the chain rule. The resulting expression for the second derivative is:

$$\begin{aligned}\frac{\partial^2 f(c)}{\partial c^2} = {} & \lambda_1\left(\frac{\lambda_2}{\lambda_1}\frac{c_2^{\,0}}{\left(c_1^{\,0}\right)^{\frac{\lambda_2}{\lambda_1}}}\left(1-\frac{\lambda_2}{\lambda_1}\right)\left(c_1\right)^{\frac{\lambda_2}{\lambda_1}-2}\right)\left(1+\frac{\lambda_2}{\lambda_1}\frac{c_2^{\,0}}{\left(c_1^{\,0}\right)^{\frac{\lambda_2}{\lambda_1}}}\left(c_1\right)^{\frac{\lambda_2}{\lambda_1}-1}\right)^{-3} - \\ & \lambda_2\left(\frac{\lambda_1}{\lambda_2}\frac{c_1^{\,0}}{\left(c_2^{\,0}\right)^{\frac{\lambda_1}{\lambda_2}}}\left(\frac{\lambda_1}{\lambda_2}-1\right)\left(c_2\right)^{\frac{\lambda_1}{\lambda_2}-2}\right)\left(1+\frac{\lambda_1}{\lambda_2}\frac{c_1^{\,0}}{\left(c_2^{\,0}\right)^{\frac{\lambda_1}{\lambda_2}}}\left(c_2\right)^{\frac{\lambda_1}{\lambda_2}-1}\right)^{-3}\end{aligned} \tag{B5}$$

When $c \to 0$, this becomes:

$$\frac{\partial f^{\,2}(0)}{\partial c^2} = 0 \tag{B6}$$

## Appendix C. Derivation of asymptotic expressions for suspension function

In this section we derive asymptotic forms of suspension functions for a binary system. We have six different asymptotic suspension functions, corresponding to different assumptions and orders of magnitudes of expansions. Here, the derivation of the suspension function is explained for the first case alone. To obtain the rest of them, the same mathematical procedure must be followed.

The assumption comprises small concentration of the first population $c_1^{\,0} = \varepsilon$ and first-order expansion.
For a binary system, the system of equations that needs to be solved is presented in Eq. ( 12) and Eq. ( 13). Using asymptotic expansions up to first order for each suspension concentration, we have:

$$c_1 = x_0 + \varepsilon x_1,\;\; c_2 = y_0 + \varepsilon y_1 \tag{C1}$$

Substituting Eqs. (C1) into Eqs. ( 12) and ( 13) we obtain Eqs. (C2) and (C3):

$$x_0 + \varepsilon x_1 + y_0 + \varepsilon y_1 = c \tag{C2}$$

$$\frac{x_0 + \varepsilon x_1}{c_1^{\,0}} = \left(\frac{y_0 + \varepsilon y_1}{c_2^{\,0}}\right)^{\frac{\lambda_1}{\lambda_2}} \tag{C3}$$

For $c_1^{\,0} = \varepsilon$, we obtain:

$$\frac{x_0 + \varepsilon x_1}{\varepsilon} = \left(\frac{y_0 + \varepsilon y_1}{1-\varepsilon}\right)^{\frac{\lambda_1}{\lambda_2}} \tag{C4}$$

Taylor expansion up to first order for the right-hand side of Eq. (C4) gives us:

$$\frac{x_0 + \varepsilon x_1}{\varepsilon} = y_0^{\frac{\lambda_1}{\lambda_2}} + \varepsilon \frac{\lambda_1}{\lambda_2} y_0^{\frac{\lambda_1}{\lambda_2}-1}\left(y_0 + y_1\right) \tag{C5}$$

Rearranging Eq. (C5) yields:

$$x_0 + \varepsilon x_1 = \varepsilon y_0^{\frac{\lambda_1}{\lambda_2}} + \varepsilon^2 \frac{\lambda_1}{\lambda_2} y_0^{\frac{\lambda_1}{\lambda_2}-1} (y_0 + y_1) \tag{C6}$$

Two Eqs. (C2) and (C6) contain several terms multiplied by powers of the small parameter $\varepsilon$ . For the equations to be valid for all $\varepsilon$ , the coefficients of each power must each be equal to zero. Considering the zeroth order coefficient in Eq. (C6) we obtain that:

$$x_0 = 0 \tag{C7}$$

which combined with the result that the zeroth order coefficients of Eq. (C2) are also zero, results in:

$$y_0 = c \tag{C8}$$

Similarly, setting the coefficient of first order powers of $\varepsilon$ to zero in Eq. (C6), results in:

$$x_1 = c^{\frac{\lambda_1}{\lambda_2}} \tag{C9}$$

and performing the same for Eq. (C2) gives:

$$y_1 = -c^{\frac{\lambda_1}{\lambda_2}} \tag{C10}$$

Substituting Eqs.((C7)-(C10)) into Eq. (C1) we obtain:

$$c_1 = \varepsilon c^{\frac{\lambda_1}{\lambda_2}},\ c_2 = c - \varepsilon c^{\frac{\lambda_1}{\lambda_2}} \tag{C11}$$

Therefore, the expression for suspension function *f(c)* is:

$$f(c) = \lambda_1 (\varepsilon c^{\frac{\lambda_1}{\lambda_2}}) + \lambda_2 (c - \varepsilon c^{\frac{\lambda_1}{\lambda_2}}) \tag{C12}$$

The same procedure has been completed for $\varepsilon=c_2^0$ and $\varepsilon=\theta_\lambda$ and the asymptotic expansions can be found in Eqs. ( 18)-( 22). The procedures involve substituting these variables for $\varepsilon$ instead of $c_1^0$ in Eq. (C3) and completing all other steps as presented here.

## Appendix D. Analytical model for 1D upscaled transport

In this section we present the analytical solution for colloidal transport in porous media with non-linear suspension function, Langmuir filtration function, and piecewise constant injected concentration. The injection scheme and (*x*,*t*) space are presented in Fig. D1. For exact integration of non-linear PDEs, we use the nonlinear methods of characteristics and of flow potentials [41], [56].

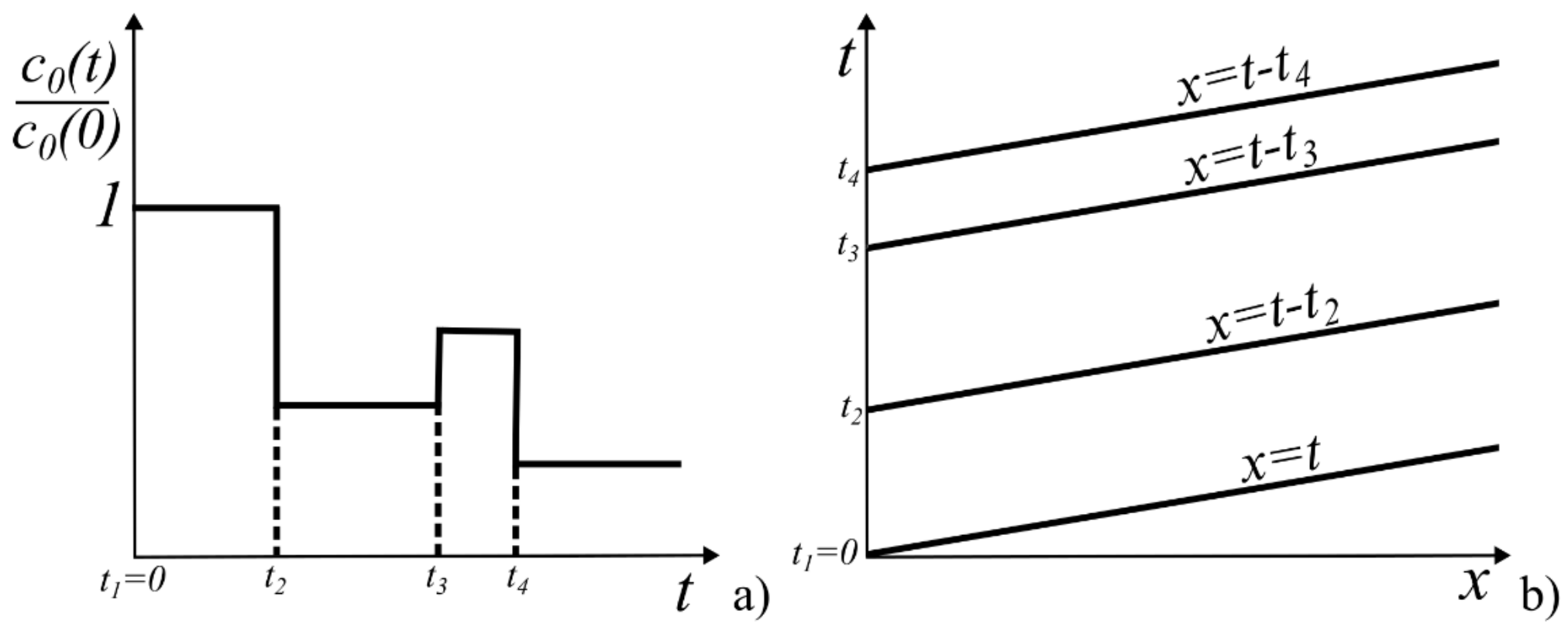

Fig. D1: Schematic for piecewise injection, a) normalised injected concentration history, b) subdivision of ($x$,$t$) phase space due to varying inlet injected concentration

Introduce a new variable:

$$\tau = t - x \tag{D1}$$

Therefore, our unknowns in Eqs. ( 8) and ( 9) are $c(x,\tau(x,t))$ and $s(x,\tau(x,t))$. Recalculating the derivatives in Eq. ( 8) using the new coordinate system, $(x,\tau)$ results in:

$$\frac{\partial c}{\partial x} = -\frac{\partial s}{\partial \tau} \tag{D2}$$

Similarly, considering Langmuir filtration function the capture rate (Eq. ( 9)) becomes:

$$\frac{\partial s}{\partial \tau} = \left(1 - \frac{s}{s_m}\right) f(c) \tag{D3}$$

From Eqs. (D2) and (D3) we obtain:

$$\frac{\partial c}{\partial x} = -\left(1 - \frac{s}{s_m}\right) f(c) \tag{D4}$$

The dimensionless initial conditions for the new system ((D2),(D3)) are:

$$\tau = -x: \; c = s = 0 \tag{D5}$$

Consider colloidal injection with piecewise concentration $c^0(t)$. The dimensionless boundary conditions are:

$$\tau = t: \; c = \frac{c^0(t)}{c^0(0)} = a_i, \;\; \forall t \in [t_i, t_{i+1}], \; t_1 = 0 \tag{D6}$$

Whereby definition of the normalised injected concentration during the first stage, $a_1$ is equal to one. Given the definition of $\tau$, we have that $\tau_i = t_i$.

Ahead of the injected particle front we have:

$$c = s = 0 \tag{D7}$$

For the first injection period, the suspension concentration behind the front can be obtained substituting $s=0$ into Eq. (D4), [41], [56]:

$$\frac{\partial c}{\partial x} = -f(c) \tag{D8}$$

$$\int_{c_1^-(x)}^{1} \frac{du}{f(u)} = x \tag{D9}$$

where $c_1^-(x)$ is the concentration immediately behind the injected particle front.

Introduce a new function:

$$\varphi(1) - \varphi\left(c^-(x,0)\right) = x, \;\; \varphi'(c) = \frac{1}{f(c)} \tag{D10}$$

Given the unit slope of the characteristics, the $(x,t)$ space is divided into regions by the changing injected concentration. Each region occupies the space between the characteristic beginning at $(0,t_i)$ and $(0,t_{i+1})$.

The following manipulations express retained concentration from equation (D4):

$$\frac{1}{f(c)}\frac{\partial c}{\partial x}+1=\frac{s}{s_m} \tag{D11}$$

$$\frac{\partial \varphi(c)}{\partial x}=\frac{1}{f(c)}\frac{\partial c}{\partial x}=\frac{\partial \varphi(c)}{\partial c}\frac{\partial c}{\partial x} \tag{D12}$$

$$s(x,\tau)=s_m\left[\frac{1}{f(c)}\frac{\partial c}{\partial x}+1\right]=s_m\left[\frac{\partial \varphi(c)}{\partial x}+1\right] \tag{D13}$$

Substituting Eq. (D13) into Eq. (D2):

$$\frac{\partial}{\partial \tau}\left(s_m\frac{\partial \varphi(c)}{\partial x}\right)+\frac{\partial c}{\partial x}=0 \tag{D14}$$

Changing the order of derivatives and integrating in $x$, accounting for boundary conditions:

$$s_m\frac{\partial \varphi(c)}{\partial \tau}+c=a_i \tag{D15}$$

$$\frac{\partial c}{(a_i-c)\,f(c)}=\frac{\partial \tau}{s_m} \tag{D16}$$

Taking derivative of both sides of Eq. (D16):

$$\int_{c_i^-(x)}^{c}\frac{du}{(a_i-u)\,f(u)}=\frac{(\tau-\tau_i)}{s_m} \tag{D17}$$

Where $c_i^-(x)$ is the concentration on the lower boundary of the region corresponding to the normalised injected concentration $a_i$ ($t+t_i<x<t+t_{i+1}$). This can be found by substituting $\tau=t_i$ into the solution for the previous region. The value for $c_1^-(x)$ is found from solving Eq. (D9).

Taking derivative in $x$ of both sides of Eq. (D17) leads to:

$$\frac{1}{(a_i-c)\,f(c)}\frac{\partial c}{\partial x}-\frac{1}{\left(a_i-c_i^-(x,\tau_i)\right)f\left(c_i^-(x,\tau_i)\right)}\frac{dc_i^-(x,\tau_i)}{dx}=0 \tag{D18}$$

From substituting Eq. (D13) into Eq. (D18) we can determine $s(x,\tau)$.

$$\frac{s(x,\tau)-s_m}{\left(a_i-c(x,\tau)\right)}=\frac{s(x,\tau_i)-s_m}{\left(a_i-c(x,0)\right)} \tag{D19}$$

Here in the first region, the initial condition is: $s(x,0)=0$. Rearranging and returning to $(x,t)$ coordinates yields:

$$s(x,t)=s_m+\left(\frac{a_i-c(x,t)}{a_i-c_i^-(x,x+t_i)}\right) \tag{D20}$$

Expressing pressure gradient in Darcy's law (Eq. ( 5)) and integrating along the core results in:

$$\Delta p(t) = 1 + \beta\phi c^0 \int_0^1 s(u,t)\,du \tag{D21}$$

The dimensionless pressure drop can then be determined as:

$$J(t) = \frac{\Delta p(t)}{\Delta p(t=0)} \tag{D22}$$